\documentclass{article}

\usepackage[english]{babel}

\usepackage[letterpaper,top=2cm,bottom=2cm,left=3cm,right=3cm,marginparwidth=1.75cm]{geometry}

\usepackage{amsmath}
\usepackage{amsfonts}
\usepackage{graphicx}
\usepackage[colorlinks=true, allcolors=blue]{hyperref}
\usepackage{subcaption}
\usepackage{natbib}
\usepackage{multirow}

\usepackage{eqnarray}

\newcommand\numberthis{\addtocounter{equation}{1}\tag{\theequation}}

\usepackage{booktabs} 

\usepackage[ruled,vlined,linesnumbered]{algorithm2e}

\usepackage{balance}
\usepackage{flushend}
\usepackage{soul}

\usepackage{subcaption}

\SetKwInput{KwInput}{Input}                
\SetKwInput{KwOutput}{Output}              
\SetAlFnt{\small}
\SetAlCapFnt{\small}
\SetAlCapNameFnt{\small}
\SetAlCapHSkip{0pt}

\newcommand{\abs}[1]{\lvert#1\rvert}

\newcommand{\R}{\mathbb{R}}

\newcommand\norm[1]{\left\lVert#1\right\rVert}

\renewcommand{\dot}[2]{\left\langle#1,#2\right\rangle}

\renewcommand{\div}[1]{\mbox{div}\,#1}
\newcommand{\grad}[1]{\nabla#1}
\newcommand{\hess}[1]{\mbox{\textbf{H}} #1}
\newcommand{\jac}[1]{\mbox{\textbf{J}} #1}

\newcommand{\cdott}{\boldsymbol{\cdot}}

\renewcommand{\L}{\mathcal{L}}

\newcommand{\f}{f}

\title{Exploring Differential Geometry in Neural Implicits}

\author{Tiago Novello, Guilherme Schardong,  Luiz Schirmer, \\
Vinicius da Silva, Helio Lopes, Luiz Velho}

\begin{document}
\maketitle

\begin{abstract}
We introduce a neural implicit framework that exploits the differentiable properties of neural networks and the discrete geometry of point-sampled surfaces to approximate them as the level sets of \textit{neural implicit functions}.

To train a neural implicit function, we propose a loss functional that approximates a signed distance function, and allows terms with high-order derivatives, such as the alignment between the principal directions of curvature, to learn more geometric details.
During training, we consider a non-uniform sampling strategy based on the curvatures of the point-sampled surface to prioritize points with more geometric details. This sampling implies faster learning while preserving geometric~accuracy when compared with previous approaches.

We also use the analytical derivatives of a neural implicit function to estimate the differential measures of the underlying point-sampled surface.
\end{abstract}

\section{Introduction}\label{s-introduction}
Level sets of neural networks have been used to represent implicit surfaces in $\R^3$ with accurate results.
In this context, the \textit{neural implicit problem} is the task of training the parameters $\theta$ of a neural network $f_\theta:\R^3\to \R$ such that its \textit{ zero-level set} $\f_\theta^{-1}(0)=\{p \,|\, \f_\theta(p)=0\}$ approximates a desired surface in $\R^3$.
We say that $f_\theta$ is a \textit{neural implicit function} and that $f_\theta^{-1}(0)$ is a \textit{neural implicit surface}.

In this work, we propose a framework to solve the neural implicit problem. The \textit{input} is a discrete sample of points from a \textit{ground truth} surface $S$, and the \textit{output} is a neural implicit function $f_\theta$ approximating the \textit{signed distance function} (SDF) of $S$.
The framework explores the \textit{differential geometry} of implicit surfaces in the learning process of $f_\theta$.
Thus, for simplicity, we consider $f_\theta$ to be a smooth function.
\textit{Sinusoidal representation networks} (SIREN)~\citep{sitzmann2020implicit} and \textit{implicit geometric regularization} (IGR)~\citep{gropp2020implicit} are examples of smooth neural networks.
We adopt as a basis the SIREN architecture which has important properties that are suitable for reconstructing signals.

Specifically,
let $\{p_i,N_i\}$ be the input composed of a sample of points and normals from a possibly \textit{unknown} surface $S$. We look for a set of parameters $\theta$ such that $f_\theta$ approximates the SDF of $S$.
Since SDFs have unit gradient we ask $f_\theta$ to satisfy the \textit{Eikonal equation} $\norm{\grad{f_\theta}}=1$ in $\{p_i\}$. Moreover, it is required the conditions $f_\theta(p_i)=0$ and $\dot{\grad{f_\theta}(p_i)}{N_i}=1$, which force the zero-level set of $f_\theta$ to interpolate $\{p_i\}$ and the gradient to be aligned to the normals $N_i$.
Additionally, to avoid spurious components in $f_\theta^{-1}(0)$, we extend these constraints to off-surface points using a SDF approximation.

The above constraints require two degrees of differentiability of $f_\theta$.
We also explore the ``alignments" between the \textit{shape operator}
$d{\frac{\grad{f_\theta}}{\norm{\grad{f_\theta}}}}$ of the neural surface $f_\theta^{-1}(0)$ and the shape operator $dN$ of $S$.
This requires one more degree of differentiability of $f_\theta$.
As the shape operators carry the intrinsic and extrinsic geometry of their surfaces, asking their alignment would require more consistency between the geometrical features of $f_\theta^{-1}(0)$ and~$S$.

In practice, we have a sample of points of $S$. Suppose that the shape operator is known at these points.
During the training of $f_\theta$, it is common to sample batches of points uniformly. However, there is a high probability of selecting points with poor geometric details on their neighborhoods, which leads to slow learning of the network.

This work proposes a sampling based on the curvature to select points with important geometric details during the training.

With a trained neural implicit function $f_\theta$ in hand, we can analytically calculate the differential geometry formulas of the corresponding neural implicit surface since we have its shape operator $d{\frac{\grad{f_\theta}}{\norm{\grad{f_\theta}}}}$. We provide the formulas along with the text.

The main contribution of our work is
a global geometric representation in the continuous setting using neural networks as implicit functions.
Besides its compact representation that captures geometrical details, this model is robust for shape analysis and efficient for computation since we have its derivatives in closed form.
The contributions can be summarized as follows:
\begin{itemize}
\item A method to approximate the SDF of a surface $S$ by a network $f_\theta$. The \textit{input} of the framework is a sample of points from $S$ (the \textit{ground truth}) endowed with its normals and curvatures, and the SDF approximation $f_\theta$ is the \textit{output}.

\item A \textit{loss functional} that allows the exploration of tools from continuous differential geometry during the training of the neural implicit function. This provides high fidelity when reconstructing geometric features of the surface and acts as an implicit regularization.

\item During the training of the network, we use the discrete differential geometry of the dataset (point-sampled surface) to \textit{sample} important regions. This provides a robust and fast training without losing geometrical~details.

\item We also use the derivatives, in closed form, of a neural implicit function to estimate the differential measures, like normals and curvatures, of the underlying point-sampled surface. This is possible since it lies in a neighborhood of the network zero-level set.

\end{itemize}

In this work, we will focus on implicit surfaces in $\R^3$. However, the definitions and techniques that we are going to describe can be easily adapted to the context of implicit $n$-dimensional \textit{manifolds} embedded in $\R^{n+1}$. In particular, it can be extended to curves and gray-scales images (graph of 2D functions).
For neural animation of surfaces, see~\citep{novello21neuralAnimation}.

\section{Related concepts and previous works}
The research topics related to our work include implicit surface representations using neural networks, discrete and continuous differential geometry, and surface reconstruction.

\subsection{Surface representation}
\label{ss-surface_representation}
A surface $S\subset\R^3$ can be represented \textit{explicitly} using a collection of \textit{charts} (\textit{atlas}) that covers $S$ or \textit{implicitly} using an implicit function that has $S$ as its zero-level set. The \textit{implicit function theorem} defines a bridge between these representations.
Consider $S$ to be a smooth surface, i.e. there is a smooth function $f$ having $S$ as its zero-level set and $\grad{f}\neq 0 $ on it. The normalized gradient $N=\frac{\grad{f}}{\norm{\grad{f}}}$ is the \textit{normal} field of $S$. The differential of $N$ is the \textit{shape operator} and gives the curvatures~of~$S$.

\vspace{0.1cm}
In practice, we usually have a point cloud $\{p_i\}$ collected from a \textit{real-world} surface $S$ whose representation is unknown.
Thus, it is common to add a structure on $\{p_i\}$ in order to operate it as a surface, for example, to compute its normals and curvatures.
The classical explicit approach is to reconstruct $S$ as a \textit{triangle mesh} having $\{p_i\}$ as its vertices.
It will be a piecewise linear approximation of $S$ with topological guarantees if $\{p_i\}$ satisfies a set of special properties~\citep{amenta2000simple}.

For simplicity, since the input of our method is a sample of points endowed with normals and curvatures, we \textit{consider} it to be the set of vertices of a triangle mesh. Then, we can use classical algorithms to approximate the normals and curvatures at the vertices. However, we could use only point cloud data and compute its normals and curvatures using well-established techniques in the literature~\citep{mitra2003estimating, alexa2003computing, mederos2003robust, kalogerakis2009extracting}.

\subsubsection{Discrete differential geometry}
Unfortunately, the geometry of the triangle mesh $T$ cannot be studied in the classical differentiable way, since it does not admit a continuous normal field.
However, we can define a \textit{discrete} notion of this field considering it to be constant on each triangle.
This implies a discontinuity on the edges and vertices. To overcome this, we use an average of the normals of the adjacent faces~\citep{meyer2003discrete}.
Hence, the variations of the normal field are concentrated on the edges and vertices of $T$.

The study of the discrete variations of the normals of triangle meshes is an important topic in \textit{discrete} differential geometry \citep{meyer2003discrete, cohen2003restricted, taubin1995estimating}.
Again, these variations are encoded in a \textit{discrete shape operator}. The principal directions and curvatures can be defined on the edges: one of the curvatures is zero, along the edge direction, and the other is measured across the edge and it is given by the dihedral angle between the adjacent faces.
Finally, the shape operator is estimated at the vertices by averaging the shape operators of the neighboring edges. We consider the approach of \citet{cohen2003restricted}.

The existent works try to discretize operators by mimicking a certain set of properties inherent in the continuous setting. Most often, it is not possible to discretize a smooth object such that all of the natural properties are preserved, this is the \textit{no free lunch} scenario. For instance, \citet{wardetzky2007discrete} proved that the existent discrete \textit{Laplacians} do not satisfy the properties of the continuous~Laplacian.

\vspace{0.2cm}

Given an (oriented) point cloud $\{p_i,N_i\}$ sampled from a surface $S$, we can try to reconstruct the SDF of $S$.
For this, points outside $S$ may be added to the point cloud $\{p_i\}$. After estimating the SDF on the resulting point cloud we obtain a set pairs $\{p_i, f_i\}$ of points and the approximated SDF values.

\subsection{Classic implicit surface reconstruction}
\textit{Radial basis functions} (RBF)~\citep{carr2001reconstruction}
is a classical technique that approximates the SDF $f$ from $\{p_i, f_i\}$. The RBF interpolant is given by: $s(p)=\sum\lambda_i\phi(\norm{p-p_i}),$ where the coefficients $\lambda_i\in\R$ are determined by imposing $s(p_i)=f_i$. The \textit{radial function} $\phi:\R^+\to \R$ is a real function, and $p_i$ are the centers of the radial basis function.
In order to consider the normals $\{N_i\}$, \citet{macedo2011hermite} proposed to approximate the function $f$ by a \textit{Hermite} radial basis function.
It is important to note that the RBF representation is directly dependent on the dataset, since its interpolant $s$ depends~on~$p_i$.

\textit{Poisson surface reconstruction}~\citep{kazhdan2006poisson} is another classical method widely used in computer graphics to reconstruct a surface from an oriented point cloud~$\{p_i,N_i\}$.

\vspace{0.2cm}

In this work,
a multilayer perceptron (MLP) network $f_\theta$ is used to overfit the unknown SDF. 
$\theta$ is trained using the point-sampled surface $\{p_i\}$ endowed with its normals and curvatures. A loss function is designed to fit the zero-level set of $f_\theta$ to the dataset.
We use the curvatures in the loss function to enforce the learning of more geometrical detail, and during the sampling to consider minibatches biased by the curvature of the data.
In Section~\ref{sec:add-experiments} we show that our neural implicit representation is comparable with the RBF method making it a flexible alternative in representing implicit functions.

Both RBF and our method look for the parameters of a function such that it fits to the signed distance of a given point-sampled surface.
Thus they are related to the \textit{regression} problem.
Differently from RBF, a neural network approach provides a compact representation and is not directly dependent on the dataset, only the training of its parameters. 
The addition of constraints is straightforward, by simply adding terms to the loss function. 
Compared to RBF, adding more constraints increases the number of equations to solve at inference time, thus increasing the problem's memory requirements.

\subsection{Neural implicit representations}

In the context of implicit surface representations using networks, we can divide the methods in three categories:
1st generation models; 2nd generation models; 3rd generation~models.

The 1st generation models correspond to \emph{global} functions of the ambient space and employ as implicit model either a \emph{indicator function} or a \emph{generalized SDF}. They use a fully connected MLP network architecture. The model is learned by fitting the input data to the model. The loss function is based either on the $L_1$ or $L_2$ norm.
The seminal papers of this class appeared in 2019. They are: Occupancy Networks~\citep{mescheder}, Learned Implicit Fields~\citep{chen2019learning}, Deep SDF~\citep{park2019deepsdf}, and Deep Level Sets~\citep{m2019}.

The 2nd generation models correspond to a set of \emph{local} functions that combined together gives a representation of a function over the whole space. These models are based either on a shape algebra, such as in \textit{constructive solid geometry}, or convolutional operators.
The main works of this category appeared in 2019-2020: Local Deep Implicit Functions~\citep{genova2019}, BSP-Net~\citep{chen2020bspnet}, CvxNet~\citep{deng2020cvxnet} and Convolutional Occupancy Networks ~\citep{peng}.

The 3rd generation models correspond to true SDFs that are given by the \emph{Eikonal} equation. The model exploits in the loss function the condition $\norm{\grad{f}} = 1$.
The seminal papers of this category appeared in 2020. They are: IGR~\citep{gropp2020implicit} and SIREN~\citep{sitzmann2020implicit}.

\vspace{0.1cm}

Inspired by the 3rd generation models, we explore smooth neural networks that can represent the SDFs of surfaces. That is, the Eikonal equation is considered in the loss function which adds constraints involving derivatives of first order of the underlying function. 
One of the main advantages of neural implicit approaches is their flexibility when defining the optimization objective.
Here we use it to consider higher order derivatives (related to curvatures) of the network during its training and sampling. This strategy can be seen as an implicit regularization which favors smooth and natural zero-level set surfaces by focusing on regions with high curvatures.
The network utilized in the framework is a MLP with a smooth activation~function.

\section{Conceptualization}
\label{s-implicit_problem}

\subsection{Implicit surfaces}
\label{ss-implicit_surfaces}
In Section~\ref{ss-surface_representation} we saw that the zero-level set $f^{-1}(0)$ of a function $f:\R^3\to \R$ represents a regular surface if $\grad{f}\neq 0$ on it. However, the converse is true, i.e. for each regular surface $S$ in $\R^3$, there is a function $f:\R^3\to \R$ having $S$ as its zero-level set~\cite[Page~116]{manfredo2016differential}.
Therefore, given a sample of points on $ S $, we could try to construct the corresponding implicit function~$f$.

\subsubsection{Differential geometry of implicit surfaces}
Let $S$ be a surface and $f:\R^3\to \R$ be its implicit function.
The differential $dN_p:T_pS\to T_pS$ of $N=\frac{\grad{\f}}{\norm{\grad{\f}}}$ at $p\in S$ is a linear map on the tangent plane $T_pS$.
The map $dN$ is called the \textit{shape operator} of $S$ and can be expressed by:
\begin{equation}\label{e-differential_of_normal}
     dN = (I-N\cdot N^\top)\frac{\hess{\f}}{\norm{\grad{\f}}} ,
\end{equation}
where $\hess{\f}$ denotes the \textit{Hessian} of $\f$ and $I$ is the identity matrix.
Thus, the shape operator is the product of the Hessian (scaled by the gradient norm) and a linear projection along the normal.

As $dN_p$ is symmetric, the spectral theorem states that there is an orthogonal basis $\{e_1,e_2\}$ of $T_pS$ (the \textit{principal directions}) where $dN_p$ can be expressed as a diagonal $2\times 2$ matrix. The two elements of this diagonal are the \textit{principal curvatures} $\kappa_1$ and $\kappa_2$, and are obtained using $dN(e_i)=-\kappa_ie_i$.

The \textit{second fundamental form} of $S$ can be used to interpret $dN$ geometrically. It maps each point $p\in S$ to the quadratic form
$\textbf{\text{II}}_p(v)=\dot{-dN_p(v)}{v}$.
Let $\alpha$ be a curve passing through $p$ with unit tangent direction $v$. The number $\kappa_n(p)=\textbf{\text{II}}_p(v)$ is the \textit{normal curvature} of $\alpha$ at $p$.
\citet{kindlmann2003curvature}~used $\kappa_n$ to control the width of the \textit{silhouettes} of $S$ during rendering.

Restricted to the unit circle of $T_p S$, $\textbf{\text{II}}_p$ reaches a maximum and a minimum (principal curvatures).
In the frame $\{e_1,e_2\}$, $\textbf{\text{II}}_p$ can be written in the quadratic form $\textbf{\text{II}}_p(v)=x_1^2\kappa_1+x_2^2\kappa_2$ with $v=x_1e_1+x_2e_2$.
Points can be classified based on their form:
\textit{elliptic} if $\kappa_1\kappa_2>0$, \textit{hyperbolic} if $\kappa_1\kappa_2<0$, \textit{parabolic} if only one $\kappa_i$ is zero, and \textit{planar} if $\kappa_1=\kappa_2=0$.
This classification is related to the \textit{Gaussian curvature} $K=\kappa_1\kappa_2$. Elliptic points have positive curvature. At these points, the surface is similar to a dome, positioning itself on one side of its tangent plane.
Hyperbolic points have negative curvature. At such points, the surface is saddle-shaped.
Parabolic and planar points have zero~curvature.

The Gaussian curvature $K$ of $S$ can be calculated using the following formula~\citep{goldman2005curvature}.
\renewcommand{\arraystretch}{1.3}
\begin{equation}
    K = -\frac{1}{\norm{\grad{\f}}^4}\det\left[\begin{array}{c|c}
        \hess{\f} & \grad{f}\\
        \hline
        \grad{f}^\top & 0
    \end{array}\right].
\end{equation}
The \textit{mean curvature} $H=(\kappa_1+\kappa_2)/2$, is an extrinsic measure that describes the curvature of $S$.
It is the half of the \textit{trace} of $dN$ which does not depend on the choice of basis.
Expanding it results in the divergence of $N$, i.e. $2H=\div \frac{\grad{f}}{\norm{\grad{f}}}$.
Thus, if $f$ is a SDF, the mean curvature can be written using the Laplacian.

\begin{figure}[ht]
    \centering
        \includegraphics[width=0.495\columnwidth]{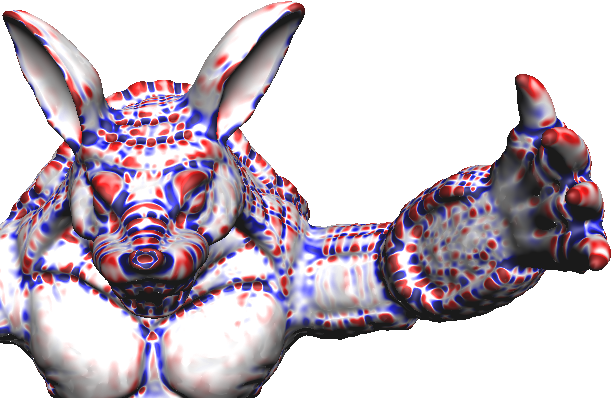}
        \includegraphics[width=0.495\columnwidth]{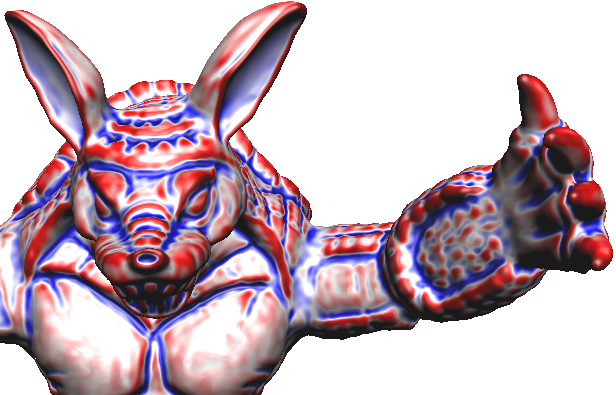}
    \caption{Gaussian and mean curvatures of the smooth Armadillo model.}
    \label{f-smooth_curvatures}
\end{figure}

An important advantage of representing a surface using level sets is that the geometric objects, like normals and curvatures, can be computed analytically --- no discretization is needed.
Figure~\ref{f-smooth_curvatures} illustrates the Gaussian and mean curvatures of a neural implicit surface approximating the Armadillo.
The corresponding network was trained using the method we are proposing.
We use the \textit{sphere tracing} algorithm~\citep{hart1996sphere} to ray cast the zero-level set.
The image was rendered using the traditional \textit{Phong} shading.
Both normal vectors and curvatures were calculated analytically using PyTorch automatic differentiation module (\texttt{torch.autograd})~\citep{NEURIPS2019_9015}. We used a transfer function to map points with high/medium/low curvatures to the red/white/blue~colors.

There are several representations of implicit functions. For example, in \textit{constructive solid geometry} the model is represented by combining simple objects using union, intersection, and difference. However, this approach has limitations, e.g. representing the Armadillo would require a prohibitive number of operations. RBF is another approach consisting of interpolating samples of the underlying implicit function which results in system of linear equation to be solved. 
A neural network is a compact option that can represent any implicit function with arbitrary precision, guaranteed by the \textit{universal approximation~theorem}~\citep{cybenko1989approximation}.

\subsection{Neural implicit surfaces}

A \textit{neural} implicit function $f_\theta:\R^3\to \R$ is an implicit function modeled by a neural network.
We call the zero-level set $f_\theta^{-1}(0)$ a \textit{neural implicit surface}.
Let $S$ be a compact surface in $\R^3$,
to compute the parameter set of $f_\theta$ such that $f_\theta^{-1}(0)$ approximates $S$, it is common to consider the \textit{Eikonal} problem:
\begin{align}\label{e-3d_eikonal_equation}
\begin{cases}
\norm{\nabla f_\theta}=1 & \text{  in } \mathbb{R}^3,\\
f_\theta =0 & \text{  on } S.
\end{cases}
\end{align}
The \textit{Eikonal equation} $\norm{\nabla f_\theta}=1$ asks for $f_\theta$ to be a \textit{SDF}. The \textit{Dirichlet condition}, $f_\theta=0$ on $S$, requires $f_\theta$ to be the signed distance of a set that contains $S$. These two constraints imply the \textit{Neumann condition}, $\frac{\partial f_\theta}{\partial N}=1$ on $S$.
Since $\frac{\partial f_\theta}{\partial N}=\dot{\nabla f_\theta}{N}$, Neumann constraint forces $\nabla f_\theta$ to be aligned to the normal field~$N$.
These constraints require two degree of differentiability of $f_\theta$,
thus, we restrict our study to smooth networks.

There are several advantages of using neural surfaces.
Besides having the entire framework of neural networks available, these functions have a high capacity of representation.
We also have access to the differential geometry tools of neural surfaces, for this, we only need the Hessian and gradient operators of the network since these are the ingredients of the shape operator (Eq.~\ref{e-differential_of_normal}).
As a consequence, we can design loss functions using high-order differential terms computed analytically.

\subsection{Learning a neural implicit surface}
\label{ss-learning_neural_implicit_surfaces}
Let $S$ be a compact surface in $\R^3$ and $\f:\R^3\to\R$ be its SDF.
Let $\f_\theta:\R^3\to\R$ be an \textit{unknown} neural implicit function.
To train $\theta$, we seek a minimum of the following loss function, which forces $f_\theta$ to be a solution of Equation~\eqref{e-3d_eikonal_equation}.
\begin{equation}\label{e-loss_function}
    \mathcal{L}(\theta)=\!\!\!\underbrace{\int\limits_{\R^3}\!\!\big|1-\norm{\grad{\f_{\theta}}}\big|dp}_{\mathcal{L}_{\text{Eikonal}}} +\!\! \underbrace{\int\limits_{S}\abs{\f_{\theta}}dS}_{\mathcal{L}_{\text{Dirichlet}}}+\!\!\underbrace{\int\limits_{S}\!\!1-\dot{\frac{\grad{\f_{\theta}}}{\norm{\grad{\f_{\theta}}}}}{N}dS}_{\mathcal{L}_{\text{Neumann}}}.
\end{equation}
${\mathcal{L}_{\text{Eikonal}}}$ encourages $\f_{\theta}$ to be the SDF of a set $\mathcal{X}$ by forcing it to be a solution of $\norm{\grad{\f_{\theta}}}=1$. ${\mathcal{L}_{Dirichlet}}$ encourages $\mathcal{X}$ to contain $S$. ${\mathcal{L}_{\text{Neumann}}}$ asks for $\grad{\f_{\theta}}$ and the normal field of $S$ to be aligned.
It is common to consider an additional term in Equation~\eqref{e-loss_function} penalizing points outside $S$, this forces $\f_{\theta}$ to be a SDF of $S$, i.e. $\mathcal{X}=S$.
In practice, we extended ${\mathcal{L}_{Dirichlet}}$ to consider points outside $S$, for this we used an approximation of the SDF of $S$.

We investigate the use of the shape operator of $f_\theta^{-1}(0)$ to improve $\mathcal{L}$, by forcing it to align with the \textit{discrete} shape operator of the ground truth point-sampled surface.
For the sampling of points used to feed a discretization of $\mathcal{L}$, the discrete curvatures access regions containing appropriate~features.

\subsection{Discrete surfaces}
\label{ss-discrete_surfaces}
\begin{figure}[hh]
    \centering
        \includegraphics[width=0.35\columnwidth]{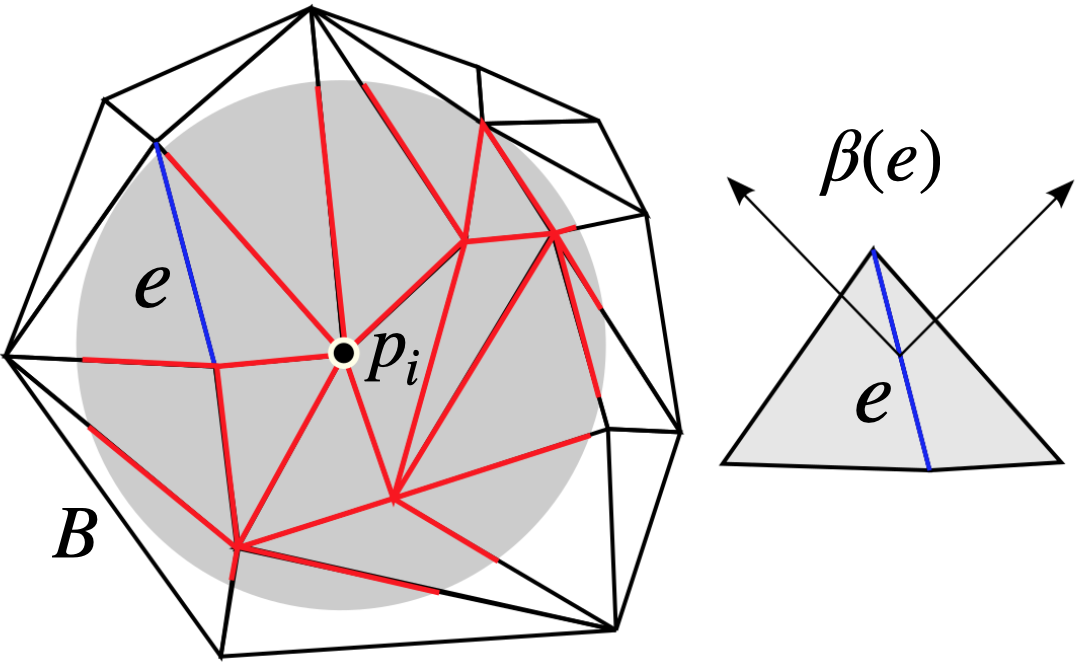}
    \caption{Discrete shape operator setting.}
    \label{f-discrete_shape_operator}
\end{figure}

Let $T=(V,E,F)$ be a triangle mesh approximating $S$. $V~=~\{p_i\}$ are the vertices, $E$ denotes the edge set, and $F$ denotes the faces.
The discrete curvatures at an edge $e$ can be estimated using $\beta(e)\,\bar{e}\cdot\bar{e}^\top$, where $\beta(e)$ is the signed dihedral angle between the two faces adjacent to $e$ and $\bar{e}$ is a unit vector aligned to $e$.
Then, the \textit{discrete shape~operator} can be defined on a vertex $p_i$ by averaging the shape operators of the neighboring edges~\citep{cohen2003restricted}.
\begin{equation}\label{e-discrete_shape_operator}
    \mathcal{S}(p_i)=\frac{1}{\text{area}(B)}\sum_{e\in E}\beta(e)\,|e\cap B|\,\bar{e}\cdot\bar{e}^\top.
\end{equation}
Where $B$ is a neighborhood of $p_i$ and $|e\cap B|$ is the length~of~$e\cap B$.
Figure~\ref{f-discrete_shape_operator} shows a schematic illustration of this operator.
It is common to consider $B$ being the dual face of $p_i$.
This operator is the discrete analogous to the shape operator~(Eq.~\eqref{e-differential_of_normal}).

Again, the discrete shape operator is a $3\times 3$ matrix, and in this case it is symmetric: there are exactly three eigenvalues and their respective eigenvectors. The normal $N_i$ is the eigenvector associated with the smaller (in absolute) eigenvalue. The remaining eigenvalues are the principal curvatures, and their associated eigenvectors are the principal directions. The principal curvatures and principal directions are permuted.
The \textit{discrete Gaussian} curvature $K_i$ and \textit{mean} curvature $H_i$ at a vertex $p_i$ are the product and the average of the principal discrete curvatures.

\section{Differentiable Neural Implicits}
\label{s-Diff_neural_implicits}
This section explores the differential geometry of the level sets of networks during their training. For the sampling, we use the curvatures of the dataset to prioritize important features.

\subsection{Neural implicit function architecture}\label{ss-SIREN}
We consider the neural function $f_\theta:\R^3\to \R$ to be defined~by
\begin{equation}\label{e-network-architecture}
    \f_\theta(p)=W_n\circ f_{n-1}\circ f_{n-2}\circ \cdots \circ f_{0}(p)+b_n
\end{equation}
\noindent
where $f_{i}(p_i)=\varphi(W_i p_i + b_i)$ is the $i$th layer, and $p_i$ is the output of $f_{i-1}$, i.e. $p_i=f_{i-1}\circ \cdots \circ f_{0}(p)$. The smooth \textit{activation function} $\varphi:\R\to \R$ is applied to each coordinate of the affine map given by the linear map $W_i:\R^{N_i}\to \R^{N_{i+1}}$ translated by the \textit{bias} $b_i\in\R^{N_{i+1}}$. The linear operators $W_i$ can be represented as matrices and $b_i$ as vectors. Therefore, the union of their coefficients corresponds to the parameters $\theta$ of $\f_\theta$.

We consider $\varphi$ to be the sine function since the resulting network is suitable for reconstructing signals and can approximate all continuous functions in the cube $[-1,1]^3$~\citep{cybenko1989approximation}. Recently, \citet{sitzmann2020implicit} proposed an initialization scheme for training the parameters of the network in the general context of signal reconstruction --- implicit surfaces being a particular case.
Here, we explore the main properties of this definition in implicit surface reconstruction.
For example, its first layer is related to a \textit{Fourier feature mapping}~\citep{benbarka2021seeing}, which allows us to represent high-frequency three-dimensional implicit functions.

Another property of this network is its smoothness, which enables the use of differential geometry in the framework.
For example, by Equation~\ref{e-differential_of_normal}, the shape operator of a neural surface can be computed using its gradient and Hessian. These operators are also helpful during training and shading.
As a matter of completeness we present their formulas~below.

\subsubsection{Gradient of a neural implicit function}\label{sss-SIREN_smooth}
The neural implicit function $f_{\theta}$ is smooth since its partial derivatives (of all orders) exist and are continuous. Indeed, each function $f_i$ has all the partial derivatives because, by definition, it is an affine map with the smooth activation function $\varphi$ applied to each coordinate. Therefore, the chain rule implies the smoothness of $f_{\theta}$.
We can compute the gradient of $f_{\theta}$ explicitly:
\begin{align*}
    \grad{\f_\theta}(p) &=\jac W_n\big( f_{n-1}\circ \cdots \circ f_{0}(p)\big) \cdott \cdots \cdott \jac f_1\big(f_0(p)\big)\cdott \jac f_0(p)\\
    &=W_n\cdott \jac f_{n-1} (p_{n-1})\cdott \cdots \cdott \jac f_1(p_{1})\cdott \jac f_0(p)\numberthis\label{e-neural_implicit_gradient},
\end{align*}
where $\jac$ is the \textit{Jacobian}
and $p_i=f_{i-1}\circ \cdots \circ f_{0}(p)$. Calculations lead us to an explicit formula for the Jacobian of $f_i$ at $p_i$.
\begin{align}\label{e-jacobian_layers}
    \jac f_{i}(p_i)=W_i\odot \varphi'\big[a_i|\cdots|a_i\big].
\end{align}
$\odot$ is the \textit{Hadamard} product, and the matrix $\big[a_i|\cdots |a_i\big]$ has $N_i$ copies of $a_i=W_i(p_i)+b_i\in\R^{N_{i+1}}$.

\subsubsection{Hessian of a neural implicit function}
Recall the \textit{chain rule} formula for the \textit{Hessian} operator of the composition of two maps $f:\R^m\to\R^n$ and $g:\R^n\to\R$:
\begin{align}
    \hess{(g\!\circ \!f)}(p)\!= \!\jac{f}(p)^\top\!\!\cdott \hess{g}\big(f(p)\big)\!\cdott\!\jac{f}(p) + \jac{g}\big(f(p)\big)\cdott\hess{f}(p)
\end{align}
We use this formula to compute the hessian $\hess{f_\theta}$ of the network $f_\theta$ using induction on its layers.

Let $f=f_{i-1}\circ \cdots \circ f_{0}$ be the composition of the first $i$ layers of $f_\theta$, and $g$ be the $l$-coordinate of the $i$-layer $f_{i}$.
Suppose we have the hessian $\hess{f}(p)$ and jacobian $\jac{f}(p)$, from the previous steps of the induction. Then we only have to compute the hessian $\hess{g}\big(f(p)\big)$ and jacobian $\jac{g}\big(f(p)\big)$ to obtain $ \hess{(g\circ f)}(p)$.
Equation~\eqref{e-jacobian_layers} gives the formula of the jacobian of a hidden layer.

Expanding the Hessian $\hess{g(p)}$ of the layer $g(p)=\varphi(w_lp+b_l)$ gives us the following formula.
\begin{align}
    \hess{g(p)}=w_l^\top w_l\cdot\varphi''(w_lp+b_l).
\end{align}
Where $w_l$ is the $l$-line of $W$ and $b_l$ is the $l$-coordinate of the bias~$b$.
When using $\varphi=\sin$, we have $ \hess{g(p)}=~-w_l^\top w_l\cdot g(p)$.

\subsection{Loss functional}
\label{ss-loss_functional}
Let $S$ be a compact surface and $f:\R^3\to \R$ be its SDF.
Here, we explore the loss functional $\L=\L_{\text{Eikonal}}+\L_{\text{Dirichlet}}+\L_{\text{Neumann}}$ used to train neural implicit functions.
The training consists of seeking a minimum of $\mathcal{L}$ using the \textit{gradient descent}.
We present ways of improving the Dirichlet and Neumann~constraints.

\subsubsection{Signed distance constraint}
\label{sss-SDF_constraints}
In practice we have a sample of points $\{p_i\}_{i=1}^n$ being the vertices of a triangulation $T$ of $S$.
Then we replace $\L_{\text{Dirichlet}}$ by
\begin{align}
    \label{eq:discrete-dirichlet}
    \widetilde{\L}_{\text{Dirichlet}}(\theta)=\frac{1}{n}\sum_{i=1}^n\abs{\f_{\theta}(p_i)}.
\end{align}
Equation~\ref{eq:discrete-dirichlet} forces $f_\theta=f$ on $\{p_i\}$, i.e. it asks for $\{p_i\}~\subset~f_\theta^{-1}(0)$.
However, the neural surface $f_\theta^{-1}(0)$ could contain undesired spurious components.
To avoid this, we improve $\widetilde{\L}_{\text{Dirichlet}}$ by including off-surface points.
For this, consider the point cloud $\{p_i\}_{i=1}^{n+k}$ to be the union of the $n$ vertices of $T$ and a sample of $k$ points in $\R^3-~S$. The constraint can be extended as follows.
\begin{align}
    \label{eq:discrete-dirichlet-osp}
    \widetilde{\L}_{\text{Dirichlet}}(\theta)=\frac{1}{n+k}\sum_{i=1}^{n+k}\abs{f_{\theta}(p_i)-f(p_i)}
\end{align}
The algorithm in Section~\ref{sss-SDF} approximates $f$~in~$\{p_i\}_{i=n+1}^{n+k}$.

\citet{sitzmann2020implicit} uses an additional term $\int e^{-100\abs{f_{\theta}}}dp$, to penalize off-surface points. However, this constraint takes a while to remove the spurious components in $f_\theta^{-1}(0)$.
\citet{gropp2020implicit} uses a pre-training with off-surface points. Here, we use an approximation of the SDF during the sampling to reduce the error outside the surface.
This strategy is part of our framework using computational/differential geometry.

\subsubsection{Signed distance function}
\label{sss-SDF}
Here we describe an approximation of the SDF $f$ of $S$ for use during the training of the network $\f_\theta$.
For this, we simply use the point-sampled surface consisting of $n$ points $\{p_i\}$ and their normals $\{N_i\}$ to approximate the absolute of $f$:
\begin{equation}
    |f(p)| \approx \min_{i \le n} \norm{p - p_i}
    \label{eq:closest-point}
\end{equation}

The sign of $\f(p)$ at a point $p$ is negative if $p$ is inside $S$ and positive otherwise.
Observe that for each vertex $p_i$ with a normal vector $N_i$, the sign of $\dot{p-p_i}{N_i}$ indicates the side of the tangent plane that $p$ belongs to.
Therefore, we approximate the sign of $\f(p)$ by adopting the dominant signs of the numbers $\dot{p-p_j}{N_j}$, where $\{p_j\}\subset V$ is a set of vertices close to $p$. This set can be estimated using a spatial-indexing structure such as Octrees or KD-trees, to store the points $\{p_i\}$. Alternatively, we can employ \textit{winding numbers} to calculate the sign of $\f(p)$. Recent techniques enable a fast calculation of this function and extend it to point clouds~\citep{barill2018winding}.

\subsubsection{Loss function using curvatures}

We observe that instead of using a simple loss function, with the eikonal approach, our strategy using model curvatures leads to an implicit regularization.
The on-surface constraint $\int 1-\dot{\grad{f_\theta}}{N}dS$ requires the gradient of $f_\theta$ to be aligned to the normals of $S$.
We extend this constraint by asking for the matching between the shape operators of $\f_\theta^{-1}(0)$ and $S$.
This can be achieved by requiring the alignment between their eigenvectors and the matching of their eigenvalues:
\begin{align}\label{e-shape_operator_constraint}
    \int\limits_{S}\sum_{i=1,2,3}\Big(1-\dot{(e_i)_\theta}{e_i}^2 + \abs{(\kappa_i)_\theta-\kappa_i}\Big)dS,
\end{align}
where $(e_i)_\theta$ and $(\kappa_i)_\theta$ are the eigenvectors and eigenvalues of the shape operator of $\f_\theta^{-1}(0)$, and $e_i$ and $\kappa_i$ are the eigenvectors and eigenvalues of the shape operator of $S$.
We opt for the square of the dot product because the principal directions do not consider vector orientation.
As the normal is, for both $\f_\theta^{-1}(0)$ and $S$, one of the shape operator eigenvectors associated to the zero eigenvalue, Equation~\eqref{e-shape_operator_constraint} is equivalent~to:
\begin{align}\label{e-shape_operator_constraint_1}
    \!\!\int\limits_{S}\!\!1-\!\dot{\frac{\grad{\f_{\theta}}}{\norm{\grad{\f_{\theta}}}}}{N}dS \!\!+\!\! \int\limits_{S}\!\!\!\sum_{i=1,2}\Big(\!1\!-\!\dot{(e_i)_\theta}{e_i}^2 \!+\! \abs{(\kappa_i)_\theta-\kappa_i}\Big)dS
\end{align}
The first integral in Equation~\eqref{e-shape_operator_constraint_1} coincides with $\mathcal{L}_{Neumann}$. In the second integral, the term $1-\dot{(e_i)_\theta}{e_i}^2$ requires the alignment between the principal directions, and $\abs{(\kappa_i)_\theta-\kappa_i}$ asks for the matching of the principal curvatures.
Asking for the alignment between $(e_1)_\theta$ and ${e_1}$ already forces the alignment between $(e_2)_\theta$ and ${e_2}$, since the principal directions are orthogonal.

We can weaken the second integral of Equation~\eqref{e-shape_operator_constraint_1} by considering the difference between the mean curvatures $\abs{H_\theta-H}$ instead of $\abs{(\kappa_1)_\theta-\kappa_1}+\abs{(\kappa_2)_\theta-\kappa_2}$. This is a weaker restriction because $\abs{H_\theta-H}\leq \frac{1}{2}\abs{(\kappa_1)_\theta-\kappa_1}+\frac{1}{2}\abs{(\kappa_2)_\theta-\kappa_2}$. However, it reduces the computations during optimization, since the mean curvature $H_\theta$ is calculated through the divergence of $\frac{\grad{\f_\theta}}{\norm{\grad{\f_\theta}}}$.

Next, we present the sampling strategies mentioned above for use in the training process of the neural implicit function $f_{\theta}$.

\subsection{Sampling}\label{ss-sampling}
Let $\{p_i,N_i, \mathcal{S}_i\}$ be a sample from an \textit{unknown} surface $S$, where $\{p_i\}$ are points on $S$, $\{N_i\}$ are their normals, and $\{ \mathcal{S}_i\}$ are samples of the shape operator.
$\{p_i\}$ could be the vertices of a triangle mesh and the normals and curvatures be computed using the formulas given in Section~\ref{ss-discrete_surfaces}.
Let $f_\theta:\R^3\to \R$ be a neural implicit function, as we saw in Section~\ref{ss-loss_functional}, its training consists of defining a loss functional $\mathcal{L}$ to force $f_\theta$ to be the SDF of $S$. 

In practice, $ \mathcal{L} $ is evaluated on a dataset of points dynamically sampled at training time.
This consists of a sampling of on-surface points in $\{p_i\}$ and a sampling of off-surface points in $\R^3-S$.
For the off-surface points, we opted for an uniform sampling in the domain of $ f_\theta $. Additionally, we could bias the off-surface sampling by including points in the \emph{tubular neighborhood} of $S$ --- a region around the surface given by a disjoint union of segments along the normals.

The shape operator encodes important geometric features of the data.
For example, regions containing points with higher principal curvatures $\kappa_1$ and $\kappa_2$
in absolute codify more details than points with lower absolute curvatures. These are the elliptic ($\kappa_1\kappa_2>0$), hyperbolic ($\kappa_1\kappa_2<0$), or parabolic points (when only one $\kappa_i$ is zero). Regions consisting of points close to planar, where $|\kappa_1|$ and $|\kappa_2|$ are small, contain less geometric information, thus, we do not need to visit all of them during sampling.
Also, planar points are abundant, see Figure~\ref{f-smooth_curvatures}.

We propose a non-uniform strategy to select the on-surface samples $\{p_i\}$ using their curvatures to obtain faster learning while maintaining the quality of the end result.
Specifically, we divide $\{p_i\}$ in three sets $V_1, V_2$, and $V_3$ corresponding to \textit{low}, \textit{medium}, and \textit{high} feature points.
For this, choosing $n=n_1+n_2+n_3$, with $n_i>0$ integer, and sorting $\{p_i\}$ using the \textit{feature} function $\kappa=|\kappa_1|+|\kappa_2|$, we define $V_1=\{p_i\,|\, i\leq n_1\}$, $V_2=\{p_i\,|\, i>n_1 \text{ and } i\leq n_1+n_2\}$, and $V_3=\{p_i\,|\, i>  n_1+n_2\}$.
Thus, $V_1$ is related to the planar points, and $V_2\cup V_3$ relates to the parabolic, hyperbolic and parabolic points.

Therefore, during the training of $f_\theta$, we can prioritize points with more geometrical features, those in $V_2\cup V_3$, to accelerate the learning.
For this, we sample less points in $V_1$, which contains data redundancy, and increase the sampling in $V_2$ and~$V_3$.

The partition $V=V_1\sqcup V_2 \sqcup V_3$ resembles the decomposition of the graph of an image in \textit{planar}, \textit{edge}, and \textit{corner} regions, the \textit{Harris corner detector}~\citep{harris1988combined}. Here, $V_2\cup V_3$ coincides with the union of the edge and corner regions.

We chose this partition because it showed good empirical results (see Sec~\ref{ss-exp_curvature_biased_sampling}), however, this is one of the possibilities.
Using the absolute of the Gaussian or the mean curvature as the feature function has also improved the training. In the case of the mean curvature, the low feature set contains regions close to a \textit{minimal} surface.
In future works, we intend to use the regions contained in the neighborhood of extremes of the principal curvatures, the so-called \textit{ridges} and \textit{ravines}.

\pagebreak
\section{Experiments}
We first consider the point-sampled surface of the Armadillo, with $n\!\!\!=\!\!\!172974$ vertices $\{p_i\}$, to explore the loss function and sampling schemes given in Section~\ref{s-Diff_neural_implicits}. We chose~this~mesh because it is a classic model with well-distributed curvatures.
We approximate its SDF using a network $\f_\theta$ with three hidden layers $f_i\!:\!\R^{256}\!\!\to\!\R^{256}$, each one followed by a sine activation.

We train $f_\theta$ using the loss functional $\mathcal{L}$ discussed in Section~\ref{ss-loss_functional}.
We seek a minimum of $\mathcal{L}$
by employing the ADAM \textit{algorithm}~\citep{kingma2014adam} with a \textit{learning rate} $1\text{e}-4$
using minibatches of size $2m$, with $m$ on-surface points sampled in the dataset $\{p_i\}$ and $m$ off-surface points uniformly sampled in $\mathbb{R}^3 - \{p_i\}$. After $	\big\lceil\frac{n}{m}\big\rceil$ iterations of the algorithm,
we have one \textit{epoch} of training, which is equivalent to passing through the whole dataset once.
We use the initialization of parameters of \citet{sitzmann2020implicit}.

\begin{figure}[ht]
    \centering
        \includegraphics[width=\columnwidth]{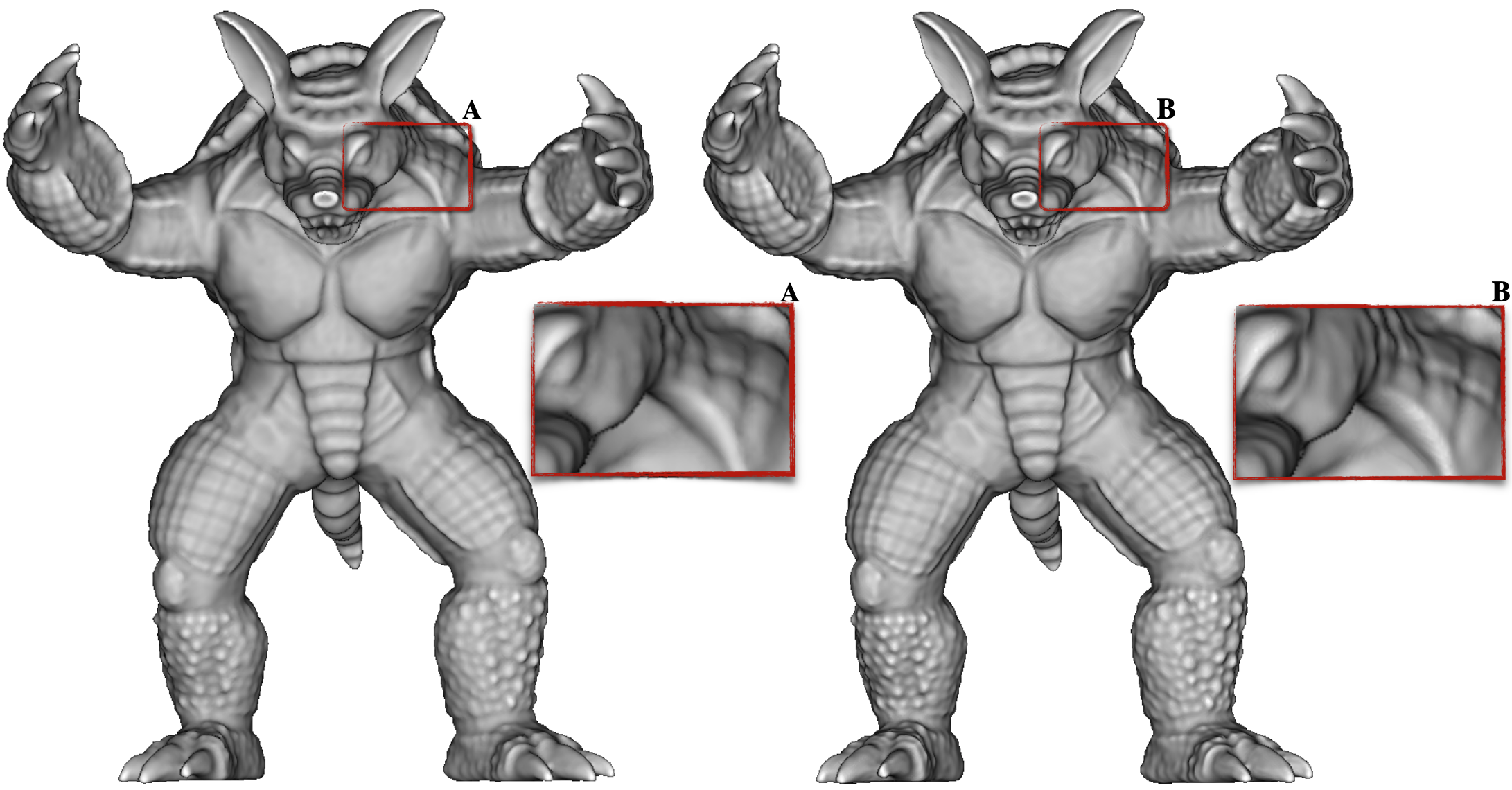}
        \caption{Comparison between the ground truth Armadillo model (right) and its reconstructed neural implicit surface (left) after $1000$ epochs of training. }
    \label{f-original_versus_reconstructed}
\end{figure}

The proposed model can represent geometric details with precision.
Figure~\ref{f-original_versus_reconstructed} shows the original Armadillo and its reconstructed neural implicit surface after $1000$ epochs of~training.

Next, we use the differential geometry of the point-sampled surface to improve the training of $f_\theta$ by adding curvature constraints in $\mathcal{L}$ and changing the sampling of minibatches in order to prioritize the points with more geometrical information.

\subsection{Loss functional}
As we saw in Section~\ref{ss-loss_functional}, we can improve the loss function $\mathcal{L}=\mathcal{L}_{\text{Eikonal}}+\mathcal{L}_{\text{Dirichlet}}+\mathcal{L}_{\text{Neumann}}$ by adding curvature terms.
Here, we use the alignment between the direction of maximum curvature $(e_1)_\theta$ of $f_\theta^{-1}(0)$ and the principal direction $e_1$ of the (ground truth) surface $S$, which leads to a regularization~term.
\begin{align}\label{eq_process}
    \mathcal{L}_{\text{Dir}}(\theta)=\int_{E}1-\dot{e_1}{(e_1)_\theta}^2dS
\end{align}
To evaluate $\mathcal{L}_{\text{Dir}}$ in $\{p_i\}$, we calculate $e_1$ in a pre-processing step considering $\{p_i\}$ be the vertices of a mesh.
Due to possible numerical errors, we restrict $\mathcal{L}_{\text{Dir}}$ to a region $E\subset S$ where $|\kappa_1-\kappa_2|$ is high. A point with $|\kappa_1-\kappa_2|$ small is close to be \textit{umbilical}, where the principal directions are not defined.

\begin{figure}[hh]
    \centering
        \includegraphics[width=0.8\columnwidth]{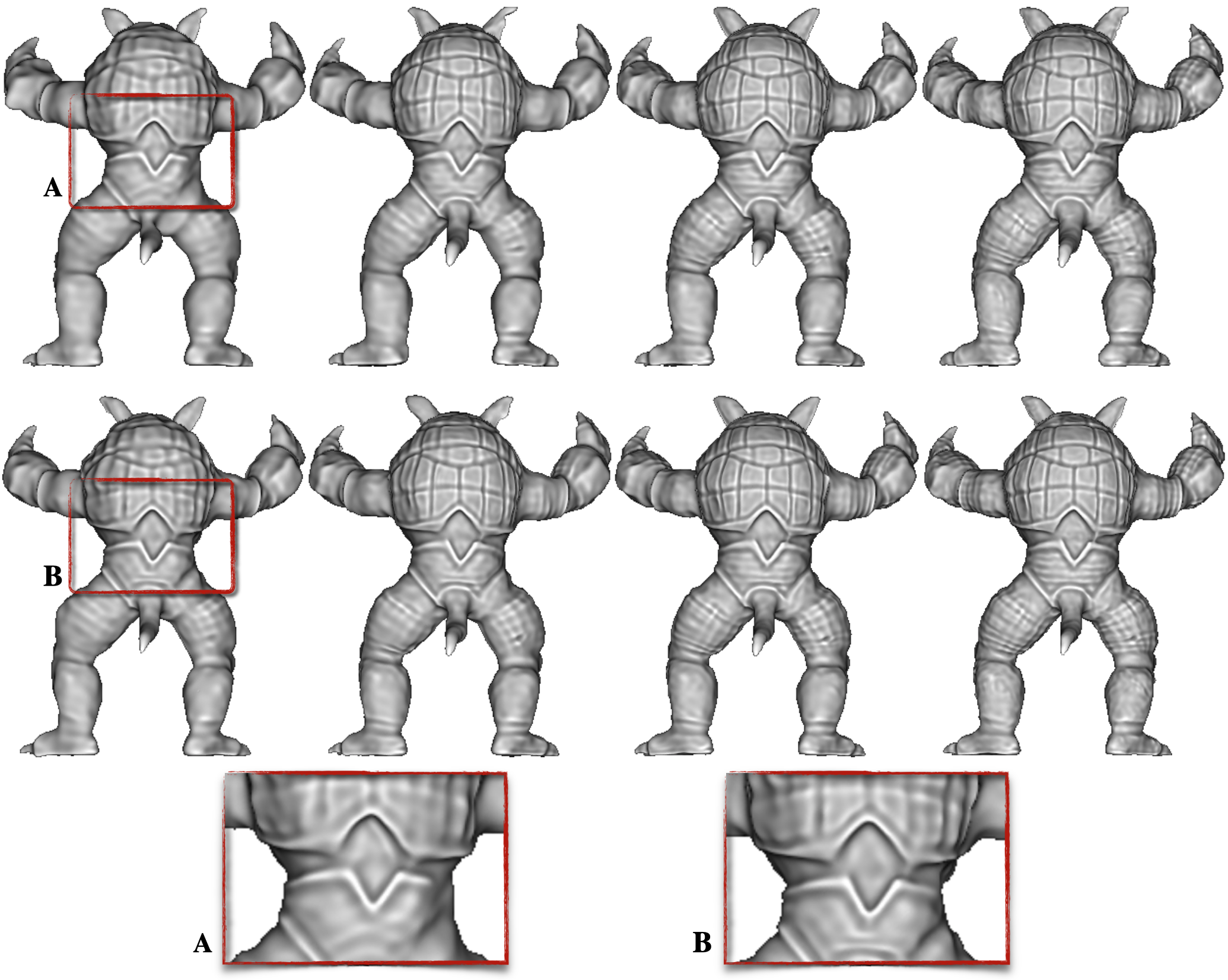}
        \caption{Neural implicit surfaces trained to approximate the Armadillo. The columns indicate the neural surfaces after $100$, $200$, $300$, and $500$ epochs of training. Line $1$ shows the results using the SDF functional. Line $2$ also consider the alignment between the principal directions in the loss functional. }
    \label{f-aligment_principal_directions}
\end{figure}

Figure~\ref{f-aligment_principal_directions} compares the training of $f_\theta$ using the loss function $\mathcal{L}$ (line~1) with the improved loss function $\mathcal{L}+\mathcal{L}_{\text{Dir}}$ (line 2).

Asking for the alignment between the principal directions during training adds a certain redundancy since we are already asking for the alignment between the normals $N$: the principal directions are extremes of $dN$. However, as we can see in Figure~\ref{f-aligment_principal_directions} it may reinforce the learning.
Furthermore, this strategy can lead to applications that rely on adding curvature terms in the loss functional. For example, we could choose regions of a neural surface and ask for an enhancement of its geometrical features. Another application could be deforming regions of a neural surface~\citep{yang2021geometry}.

\subsection{Sampling}
\label{ss-exp_curvature_biased_sampling}

This section presents experiments using the sampling discussed in Section~\ref{ss-sampling}. This prioritizes points with important features during the training implying a fast convergence while preserving quality (see Fig.~\ref{fig:loss-comparison}). Also, this analysis allows finding data redundancy, i.e., regions with similar geometry.

During training it is common to sample minibatches uniformly. Here, we use the curvatures to prioritize the important features.
Choosing $n=n_1+n_2+n_3$, with $n_i\!\in\! \mathbb{N}$, we define the sets $V_1$, $V_2$, and $V_3$ of low, medium, and high feature points.

\begin{figure}[h!]
    \centering
        \includegraphics[width=0.9\columnwidth]{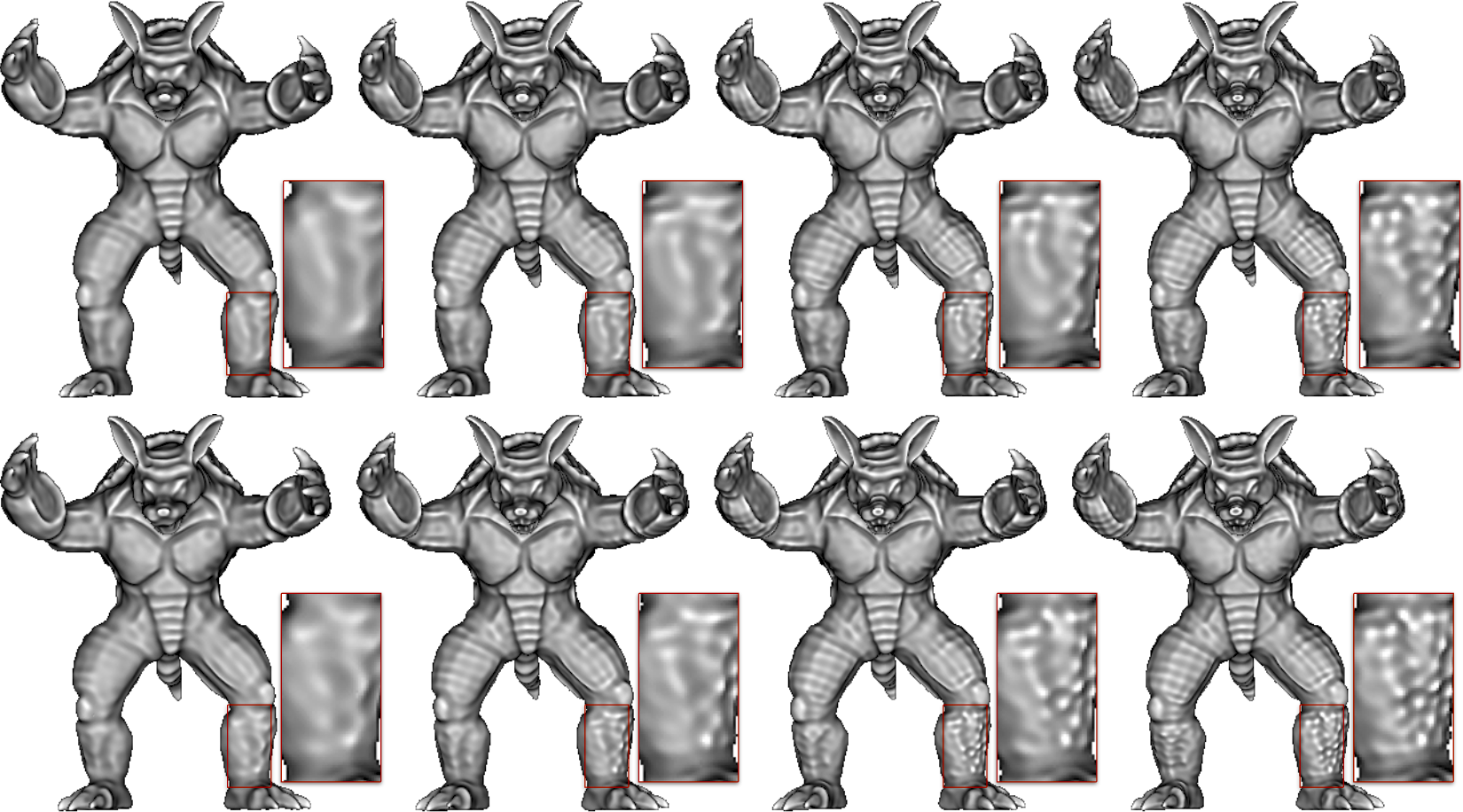}
        \caption{Neural implicit surfaces approximating the Armadillo model. The columns indicate the zero-level sets of the neural implicit functions after $29$, $52$, $76$, and $100$ epochs of training.
        Line $1$ shows the results using minibatches sampled uniformly in $V$.
        Line $2$ presents the results using the adapted sampling of minibatches with $10\%$ / $70\%$ / $20\%$ of points with low/medium/high~features.
        }
    \label{f-biased_sampling}
\end{figure}

We sample minibatches of size $m=p_1m+p_2m+p_3m=10000$, with $p_im$ points on $V_i$.
If this sampling is uniform, we would have $p_i=\frac{n_i}{n}$. Thus, to prioritize points with more geometrical features, those in $V_2$ and $V_3$, we reduce $p_1$ and increase $p_2$ and $p_3$.
Figure~\ref{f-biased_sampling} gives a comparison between the uniform sampling (first line) and the adaptive sampling (line~2) that consider $p_i=2\frac{n_i}{n}$ for $i=2,3$, i.e. it duplicates the proportion of points with medium and high features. Clearly, these new proportions depend on $n_i$. In this experiment, we use $n_1=\frac{n}{2}$, $n_2=\frac{4n}{10}$, and $n_3=\frac{n}{10}$, thus $V_1$ contains half of $V$.
This sampling strategy improved the rate convergence significantly.

Returning to minibatch sampling. In the last experiment, we were sampling more points with medium and high features in $V_2\cup V_3$ than points with low features in $V_1$. Thus the training visits $V_2\cup V_3$ more than once per epoch.
We propose to reduce the number of points sampled per epoch, prioritizing the most important ones.
For this, we reduce the size of the minibatch in order to sample each point of $V_2\cup V_3$ once per epoch.

\begin{figure}[h!]
    \centering
        \includegraphics[width=0.9\columnwidth]{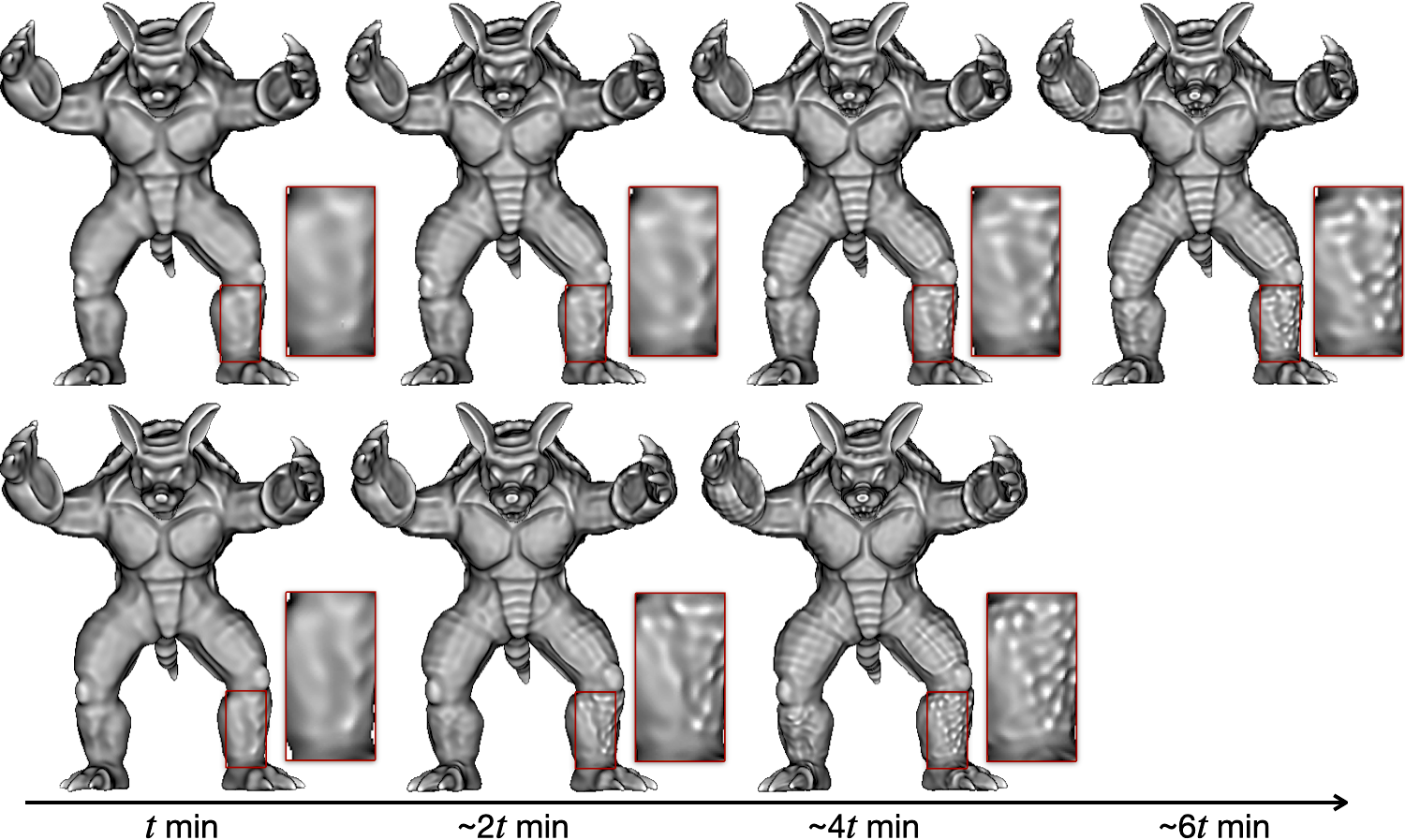}
        \caption{The columns indicate the zero-level sets of the neural implicit functions after $t$, $2t$, $4t$, and $6t$ minutes of training.
        Line $1$ shows the results using the $n$ points of the dataset per epoch and minibatches of size $m$ containing $20\%$ /$60\%$/$20\%$ of points with low/medium/high features.
        The results in line $2$ use $\frac{n}{2}$ points per epoch and minibatches of $\frac{m}{2}$ points with $20\%$ /$60\%$/$20\%$ of low/medium/high features. Both experiments use $\big\lceil\frac{n}{m}\big\rceil$ steps per~epoch.
        }
    \label{f-biased_sampling_with_less_points}
\end{figure}

Figure~\ref{f-biased_sampling_with_less_points} provides a comparison between the two training strategies. The first line coincides with the experiment presented in the second line of Figure~\ref{f-biased_sampling}.
It uses minibatches of size $m$, and $\big\lceil\frac{n}{m}\big\rceil$ iterations of the gradient descent per epoch. In the second line, we sample minibatches of size $\frac{m}{2}$ and use $\big\lceil\frac{n}{m}\big\rceil$ iterations of the gradient descent.
Then the second line visits half of the dataset per epoch.
We are using the same minibatch proportions $p_i$ and sets $V_i$, as in the previous experiment.

As we can see in Figure~\ref{f-biased_sampling_with_less_points}, going through all the points with medium and higher features once per epoch, while reducing the number of points with low features, resulted in faster training with better quality results.

\begin{figure}[ht]
    \centering
        \includegraphics[width=0.9\columnwidth]{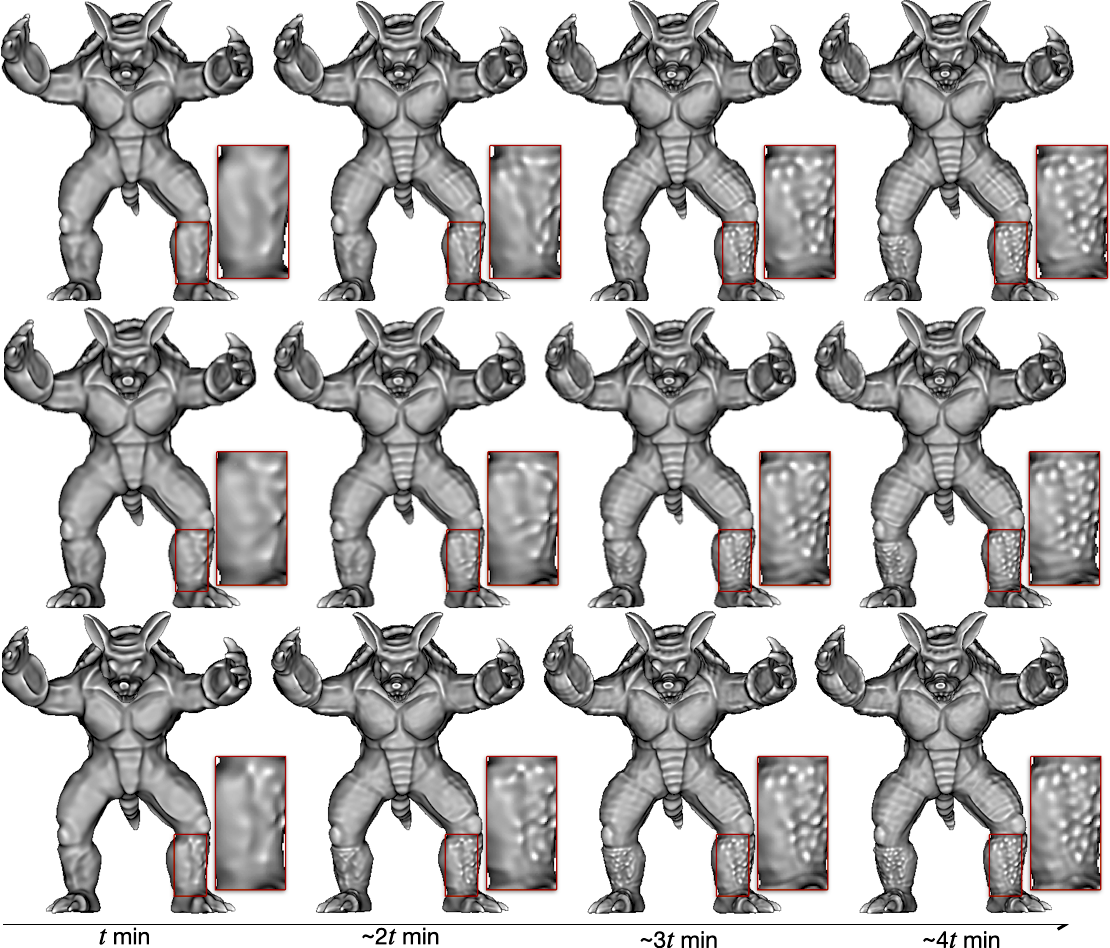}
        \caption{The columns present the zero-level sets of the model after $t$, $2t$, $3t$, and $4t$ minutes of training.
        The results in line $1$ use $\frac{n}{2}$ points per epoch and minibatches of $\frac{m}{2}$ points with $20\%$ /$60\%$/$20\%$ of low/medium/high features.
        Line $2$ shows the results using $\frac{3n}{10}$ points per epoch and minibatches of $\frac{3m}{10}$ points with $17\%$ /$67\%$/$17\%$ of low/medium/high features.
        The results in line $3$ use $\frac{n}{10}$ points per epoch and minibatches of $\frac{m}{10}$ points with $10\%$ /$70\%$/$20\%$ of low/medium/high features.
        The experiments use $\big\lceil\frac{n}{m}\big\rceil$ steps per~epoch.
        }
    \label{f-armadillo_biased_b5000-3000-1000}
\end{figure}

To keep reducing the size of the minibatches such that we visit important points once per epoch, we consider smaller medium and high feature sets. That is, to visit $V_2\cup V_3$ once per epoch it is necessary to update the sizes $n_i$ of the feature sets $V_i$.
For this, we present three experiments that use $n_1=\frac{6n}{10},\frac{75n}{100},\frac{85n}{100} $, $n_2=\frac{3n}{10}, \frac{2n}{10}, \frac{n}{10}$, and $n_3=\frac{n}{10}, \frac{5n}{100},\frac{5n}{100}$, respectively.
Then we can consider minibatches of size $\frac{m}{2}$, $\frac{3m}{10}$, and $\frac{m}{10}$, i.e. $50\%$, $30\%$ and $10\%$ of the original minibatch of size $m$.
Figure~\ref{f-armadillo_biased_b5000-3000-1000} shows the results of the experiments.
They are all using $\big\lceil\frac{n}{m}\big\rceil$ iterations per epoch, then, we visit $\frac{n}{2}$, $\frac{3n}{10}$, and $\frac{n}{10}$ points of the dataset on each epoch, respectively.
Thus, as we reduce the minibatches sizes, we remove points from $V_2\cup V_3$, which implies that we are going to learn fewer intermediate features. This can be visualized in Figure~\ref{f-armadillo_biased_b5000-3000-1000}. Observe that points with higher features, such as the shin, thighs and abdomen, are learned faster than points with lower features, mainly around the Armadillo chest.

Finding the optimal parameters of the proposed curvature-based sampling can be a hard task for a general surface. The experiments above show empirically that visiting the ``important" points once per epoch implies good geometrical results.
Another way to select important points is using the concept of \textit{ridge} and \textit{ravine} curves \citep{ohtake2004ridge}.
These are extremes of the principal curvatures along the principal directions and indicate where the surface bends sharply~\citep{belyaev1998ridges} .

\subsection{Additional experiments and comparisons}
\label{sec:add-experiments}

This section presents a comparison of our framework with RBF~\citep{carr2001reconstruction} and SIREN~\citep{sitzmann2020implicit}. For this, we introduce two analytical surfaces with ground truth in closed form: sphere and torus; and four discrete surfaces: Bunny, Dragon, Buddha and Lucy. We choose these models because they have distinct characteristics: Bunny has a simple topology/geometry, Dragon and Lucy have a simple topology with complex geometry, and Buddha has a complex topology/geometry.
It is important to mention that the goal is not showing that our method outperforms RBF, but is comparable in the function approximation~task.

For method comparison, we consider RBF since it is a well-established method to estimate an implicit function from a point-sampled surface. 
Although both RBF and our approach can approximate implicit functions, their nature is fundamentally different.
RBF solves a system of equations to weigh the influence of pairs $\{p_i, f_i\}$ on neighboring points. If we wish to include normal alignment (Hermite data) in this process~\citep{macedo2011hermite}, it demands profound changes in the interpolant estimation.
However, including Hermite data in neural implicit models demands only additional terms in the loss function, the model architecture and training process remains unchanged.

To reconstruct the sphere and the torus models we consider a network $f_\theta$ consisting of two hidden layers $f_i:\R^{80}\to\R^{80}$ and train its parameters for each model using the basic configuration given in Section~\ref{ss-loss_functional}.
We trained $f_\theta$ for $500$ epochs considering batches of $m=2500$ on-surface points and $m$ off-surface points.
For SIREN, we use the same network architecture and uniform sampling scheme, only the loss function was replaced by the original presented in~\citep{sitzmann2020implicit}.
For the RBF interpolant, we use a dataset of $m=2500$ on-surface points and $m$ off-surface~points.

We reconstruct the other models using a neural function $f_\theta$ consisting of three hidden layers $f_i:\R^{256}\to\R^{256}$.
Again, we train $\theta$ using the basic training framework.
We consider minibatches of $m=10000$ on-surface points and $m$ off-surface points.
For SIREN, we use the same network architecture and the loss function and sampling scheme given in~\citep{sitzmann2020implicit}.
For the RBF interpolant, we used a dataset of $10000$ on-surface points and $10000$ off-surface points.

Table~\ref{t-comparisons} presents the quantitative comparisons of the above experiments.
We compare the resulting SDF approximations with the ground truth SDFs using the following measures:
\begin{itemize}
    \item The absolute difference $\abs{\bar{f}-f}$, in the domain $\mathbb{R}^3 - S$ and on the surface, between the function approximation $\bar{f}$ and the ground truth SDF $f$;
    \item The normal alignment $1-\dot{\frac{\grad{\bar{f}}}{\norm{\grad{\bar{f}}}}}{\grad{f}}$ between the gradients $\grad{\bar{f}}$ and $\grad{f}$ on the surface.
\end{itemize}
We used a sample of $2500$ on-surface points, not included in the training process, to evaluate the mean and maximum values of each measure. We also ran the experiments $100$ times and took the average of the measures. Note that, for the RBF interpolation, we did not calculate the analytical derivatives because we are using a framework without support for this feature, a numerical method was employed in this case.

\renewcommand{\arraystretch}{1.2}

\begin{table}[h!]
  \centering
  \small
  \begin{tabular}{p{0.15cm}|l|l|l|l|l|l|l}
    \cline{2-8}
    \multirow{2}{*}{} & \multicolumn{1}{c|}{\multirow{2}{*}{Method}} & \multicolumn{2}{c|}{\begin{tabular}[c]{@{}c@{}}$|\bar{f}-f|$\\ in the domain\end{tabular}} & \multicolumn{2}{c|}{\begin{tabular}[c]{@{}c@{}}$|\bar{f}-f|$\\ on the surface\end{tabular}} & \multicolumn{2}{c}{\begin{tabular}[c]{@{}c@{}}Normal\\ alignment\end{tabular}} \\
    \cline{3-8}
                      & \multicolumn{1}{c|}{} & mean & max & mean & max & mean & max \\
    \hline
    \multirow{3}{*}{\rotatebox[origin=c]{90}{Sphere}} & RBF & 4e-5 & 0.021 & 5e-8 & 1e-4 & 1.81e-6 & 1.36e-5 \\
                      & SIREN & 0.129 & 1.042 & 0.0031 & 0.013 & 6e-4 & 0.005 \\
                      & Ours & 0.001 & 0.015 & 0.0018 & 0.007 & 6e-5 & 6e-4 \\
    \hline
    \multirow{3}{*}{\rotatebox[origin=c]{90}{Torus}} & RBF & 6e-4 & 0.055 & 2e-5 & 0.001 & 1.61e-5 & 3.17e-4 \\
                      & SIREN & 0.254 & 1.006 & 0.0034 & 0.013 & 0.0007 & 0.005 \\
                      & Ours & 0.003 & 0.036 & 0.0029 & 0.011 & 0.0002 & 0.002 \\
    \hline
    \multirow{3}{*}{\rotatebox[origin=c]{90}{Bunny}} & RBF & 0.002 & 0.024 & 0.0002 & 0.004 & 0.0007 & 0.453 \\
                      & SIREN & 0.145 & 0.974 & 0.0010 & 0.004 & 0.0006 & 0.019 \\
                      & Ours & 0.003 & 0.081 & 0.0015 & 0.005 & 0.0005 & 0.017 \\
    \hline
    \multirow{3}{*}{\rotatebox[origin=c]{90}{Dragon}} & RBF & 0.002 & 0.035 & 0.0006 & 0.009 & 0.0160 & 1.459 \\
                      & SIREN & 0.106 & 1.080 & 0.0010 & 0.006 & 0.0063 & 0.866 \\
                      & Ours & 0.003 & 0.104 & 0.0010 & 0.005 & 0.0034 & 0.234 \\
    \hline
    \multirow{3}{*}{\rotatebox[origin=c]{90}{Armadillo}} & RBF & 0.003 & 0.008 & 0.0030 & 0.056 & 0.0134 & 1.234 \\
                      & SIREN & 0.126 & 0.941 & 0.0010 & 0.005 & 0.0021 & 0.168 \\
                      & Ours & 0.009 & 0.136 & 0.0012 & 0.006 & 0.0016 & 0.164 \\
    \hline
    \multirow{3}{*}{\rotatebox[origin=c]{90}{Lucy}} & RBF & 0.002 & 0.048 & 0.0003 & 0.011 & 0.1581 & 1.998 \\
                      & SIREN & 0.384 & 1.048 & 0.0007 & 0.003 & 0.0070 & 0.313 \\
                      & Ours & 0.013 & 0.155 & 0.0009 & 0.006 & 0.0056 & 0.170 \\
    \hline
    \multirow{3}{*}{\rotatebox[origin=c]{90}{Buddha}} & RBF & 0.002 & 0.050 & 0.0004 & 0.010 & 0.0687 & 1.988 \\
                      & SIREN & 0.337 & 1.124 & 0.0007 & 0.008 & 0.0141 & 1.889 \\
                      & Ours & 0.096 & 0.405 & 0.0069 & 0.024 & 0.0524 & 1.967 \\
    \hline
  \end{tabular}
\caption{Comparison between RBF, SIREN, and our framework. We consider two analytical models, the sphere and the torus, and five classical computer graphics models, \textit{Bunny, Dragon, Armadillo, Happy Buddha}, and \textit{Lucy}.}
  \label{t-comparisons}
\end{table}

\begin{figure}[h!]
    \centering
    \begin{tabular}{c|c}
       \includegraphics[width=0.43\columnwidth]{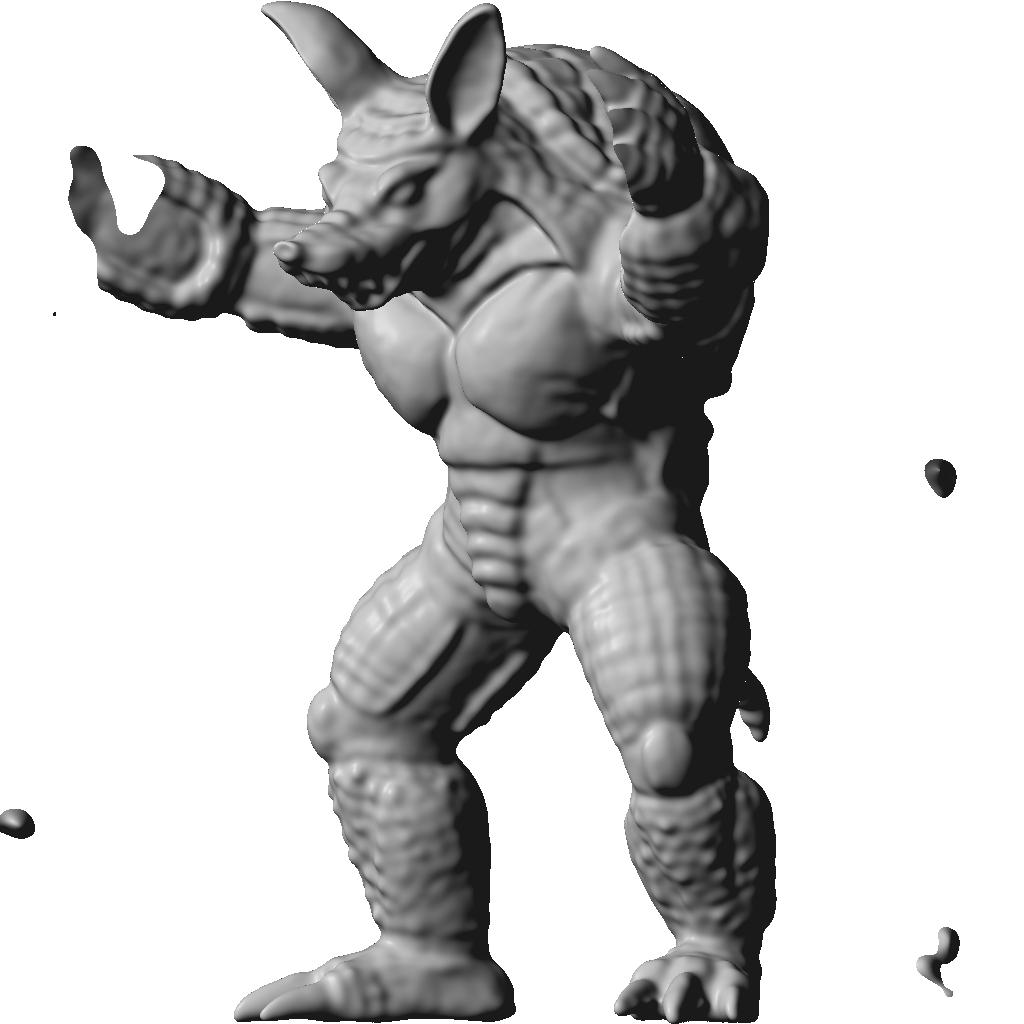}  &
       \includegraphics[width=0.43\columnwidth]{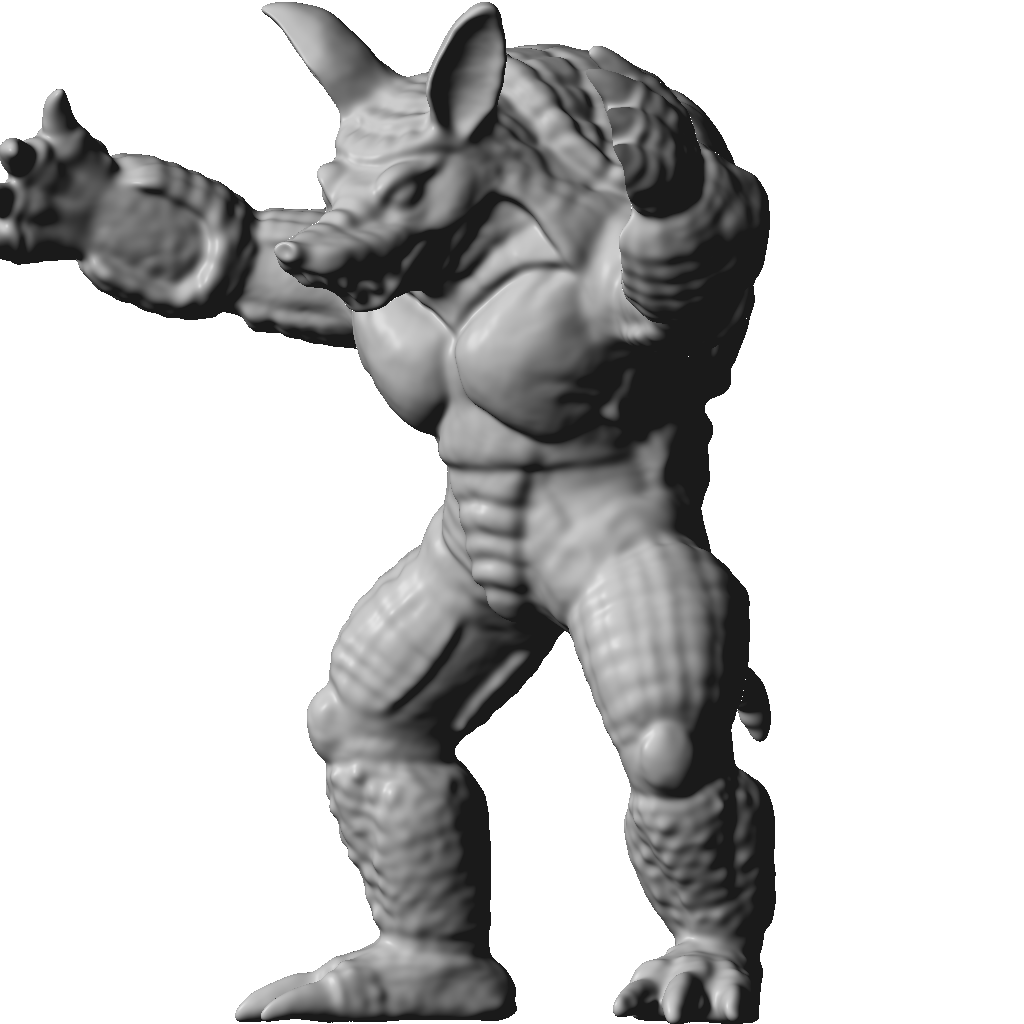}\\
       \hline
        \includegraphics[width=0.37\columnwidth]{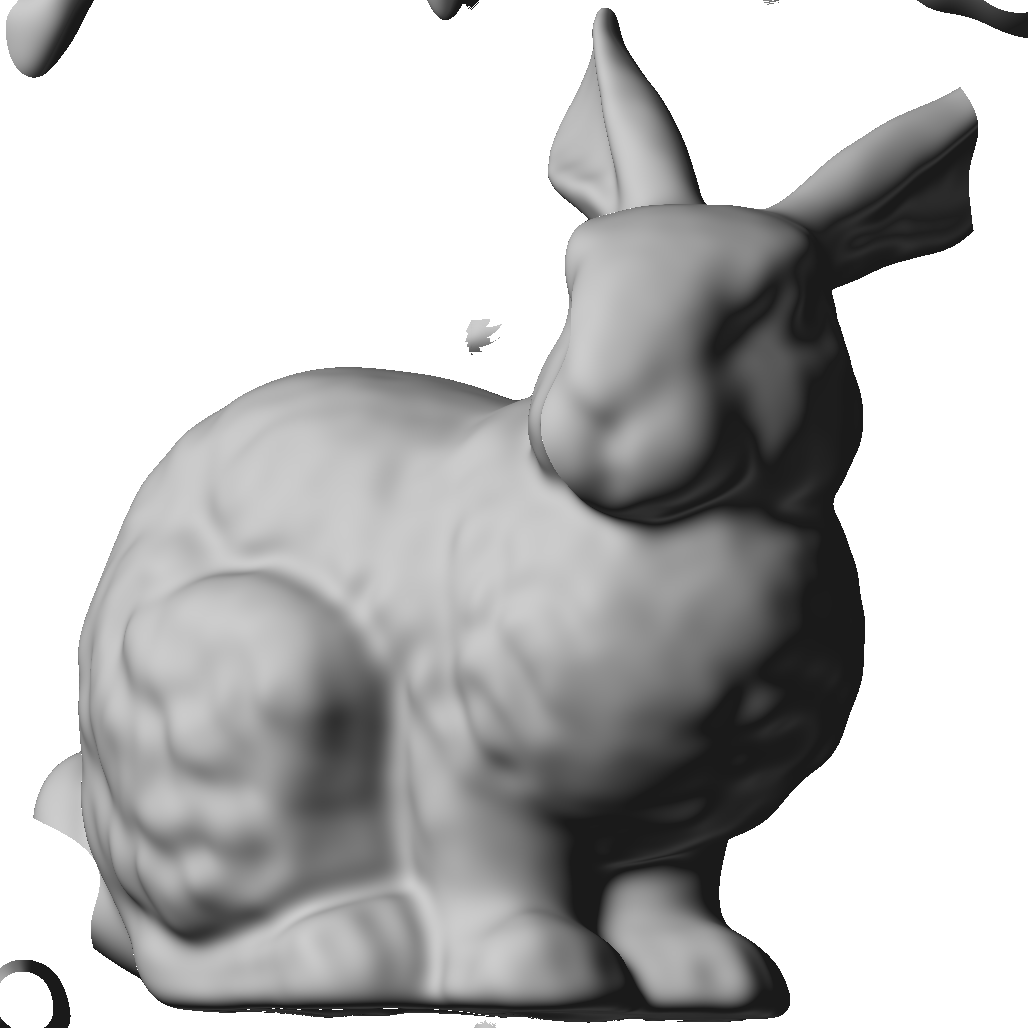}  &
       \includegraphics[width=0.37\columnwidth]{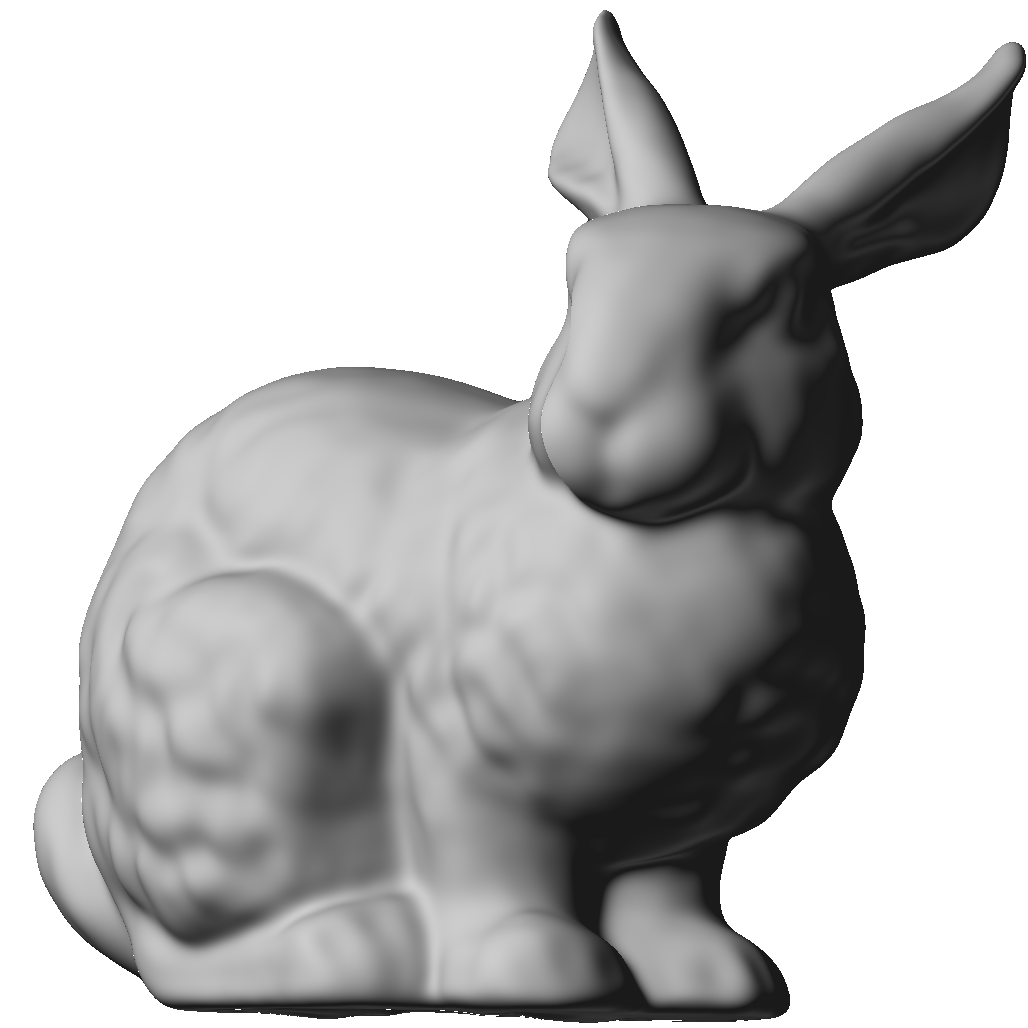}
    \end{tabular}
    \caption{Sphere tracing of neural surfaces representing the Armadillo and Bunny models. On the left, the network was trained using the SIREN framework. On the right, the results using our method. Both networks have the same architecture and were trained on the same data during $500$ epochs.}
    \label{f-sdf_comparison}
\end{figure}

Our method provides a robust SDF approximation even compared with RBF.
Figure~\ref{f-sdf_comparison} gives a visual evaluation presenting a sphere tracing of the zero-level sets of the SIREN and our method.
In both cases, we used an image resolution of $1024\!\times\! 1024$ and $80$ sphere tracing iterations.
Since we obtain a better SDF approximation the algorithm is able to ray cast the surface with precision avoiding spurious components.

We did not visualize RBF approximations because the algorithm implemented in SciPy~\citep{scipy2020}, which is employed in this work, is not fully optimized, making the ray tracing unfeasible.

Table~\ref{t-time} shows the average training and inference time for RBF, SIREN, and our method. For this experiment, we train SIREN and our method for 50 epochs using 20000 points per batch, only on CPU, to provide a fair comparison. As for RBF, we used a single batch of points to build the interpolant, with each point weighting the 300 nearest points, to diminish the algorithm's memory requirements. Training time for RBF consists mostly of creating the matrices used to solve the interpolation problem. This is a relatively simple step, thus as expected, takes only a fraction of time compared to other methods. Still regarding training time, SIREN and our method are in the same magnitude, with SIREN being slightly faster in all experiments.
This is mainly due to our method performing SDF querying at each training step. Even with efficient algorithms, this step impacts measurably in the training performance.
\begin{table}[h!]
\small
\centering
\begin{tabular}{llrr}
\hline
                           & Method & Training Time (s) & Inference Time (s) \\ \hline
\multirow{3}{*}{Bunny}     & RBF    & 0.0055            & 417.3928           \\
                           & SIREN  & 173.6430          & 0.5773             \\
                           & Ours   & 199.3146          & 0.6460             \\ \hline
\multirow{3}{*}{Dragon}    & RBF    & 0.0046            & 411.1710           \\
                           & SIREN  & 319.8439          & 0.5565             \\
                           & Ours   & 391.4102          & 0.5885             \\ \hline
\multirow{3}{*}{Armadillo} & RBF    & 0.0045            & 392.0836           \\
                           & SIREN  & 380.5361          & 0.9522             \\
                           & Ours   & 443.3634          & 0.9290             \\ \hline
\multirow{3}{*}{Buddha}    & RBF    & 0.0044            & 410.6234           \\
                           & SIREN  & 1297.0681         & 0.9158             \\
                           & Ours   & 1646.2311         & 0.9689             \\ \hline
\multirow{3}{*}{Lucy}      & RBF    & 0.0077            & 358.7987           \\
                           & SIREN  & 560.1297          & 0.8888             \\
                           & Ours   & 654.1596          & 0.8023             \\ \hline
\end{tabular}
\caption{Comparison between RBF, SIREN, and our framework. We use the same models as in Table~\ref{t-comparisons}, except for the sphere and torus.}
\label{t-time}
\end{table}


Regarding inference time, both our method and SIREN take less than a second for all models in a $64^3$ grid. As for RBF, the inference time is close to 400 seconds for all tested cases. It is affected by the size of the interpolant, which explains the proximity in inference performance even for complex models (Buddha and Dragon). 
The RBF inference could be improved using \textit{partition of unity}~\citep{ohtake:2003,ohtake:2006} or \textit{fast
multipole method}~\citep{greengard1997fast}. Here, we opt for the implementation in Scipy~\citep{scipy2020} of the RBF approach since it is widely available.

\begin{figure}[h!]
    \centering
    \includegraphics[width=0.43\columnwidth]{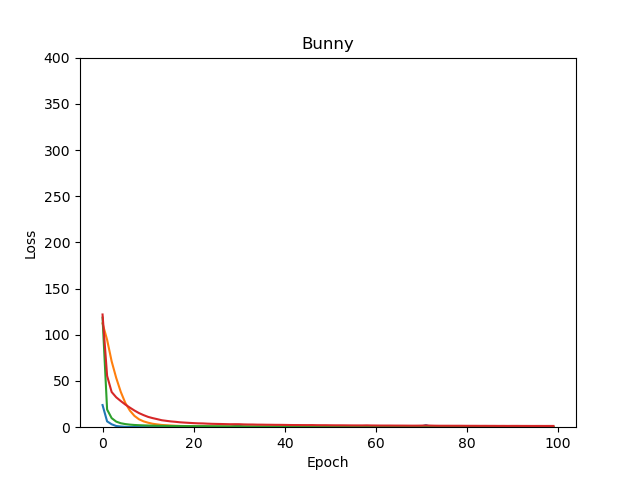}
    \includegraphics[width=0.43\columnwidth]{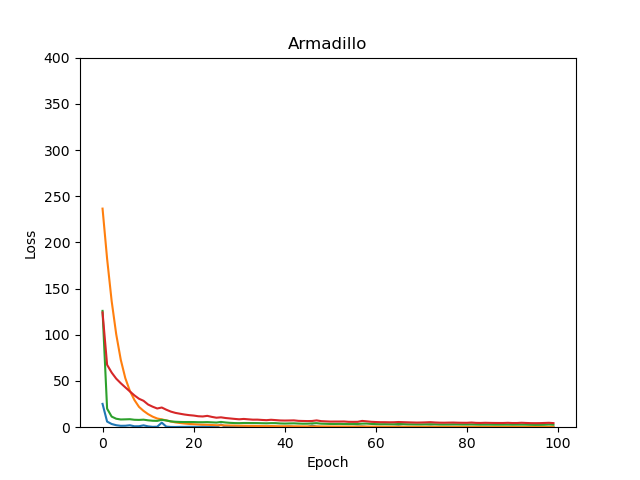}
    \includegraphics[width=0.43\columnwidth]{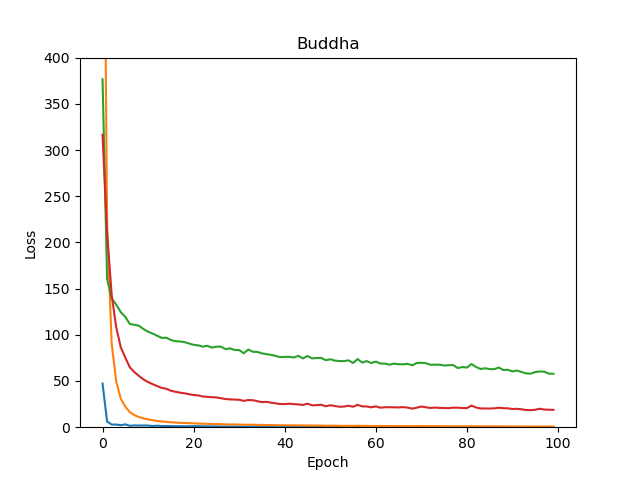}
    \includegraphics[width=0.43\columnwidth]{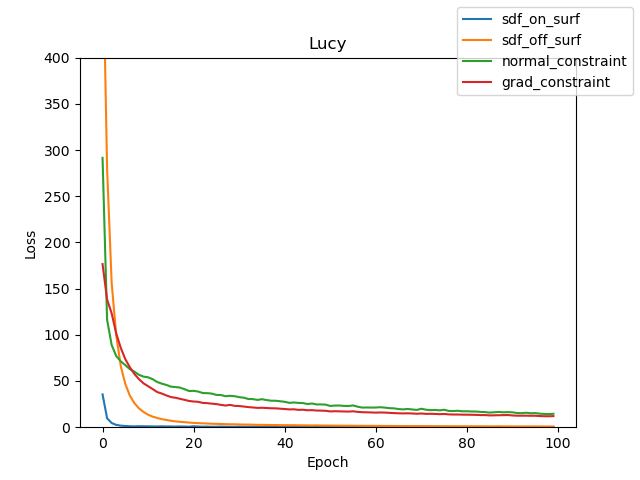}
    \caption{Training loss per epoch for all constraints for the Bunny (top-left), Armadillo (top-right), Buddha (bottom-left), and Lucy (bottom-right).}
    \label{fig:loss-comparison}
\end{figure}

Figure~\ref{fig:loss-comparison} shows the training loss per epoch for each considered model. We did not include the Dragon because its loss function behavior is similar to the Bunny. Note that the Dirichlet condition for on-surface points (sdf\_on\_surf) quickly converges and approaches zero at the first $5$ epochs. In all tested cases, the off-surface Dirichlet condition (sdf\_off\_surf) converges quickly as well, usually by the first $20$ epochs. The Eikonal/Neumann constraints take longer to converge, with the notable example of Buddha, where the Neumann constraint remains at a high level, albeit still decreasing, after $100$~epochs.

\subsection{Curvature estimation}

\begin{figure}[h!]
    \begin{subfigure}[b]{0.47\columnwidth}
         \includegraphics[width=\columnwidth]{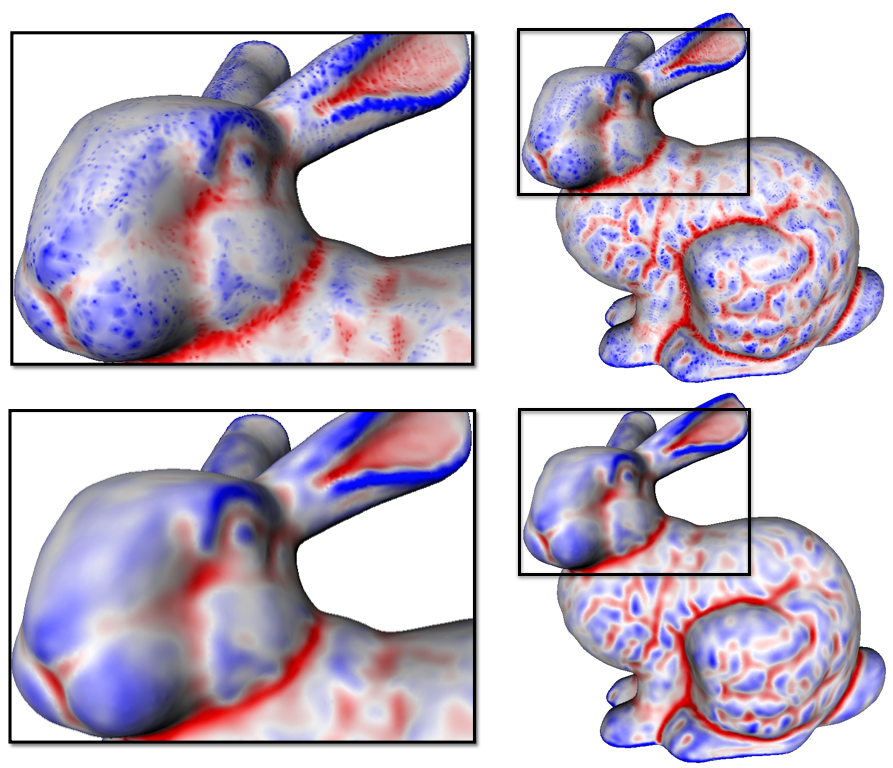}
         \caption{Bunny}
         \label{curv_bunny}
     \end{subfigure}
     \begin{subfigure}[b]{0.52\columnwidth}
         \includegraphics[width=\columnwidth]{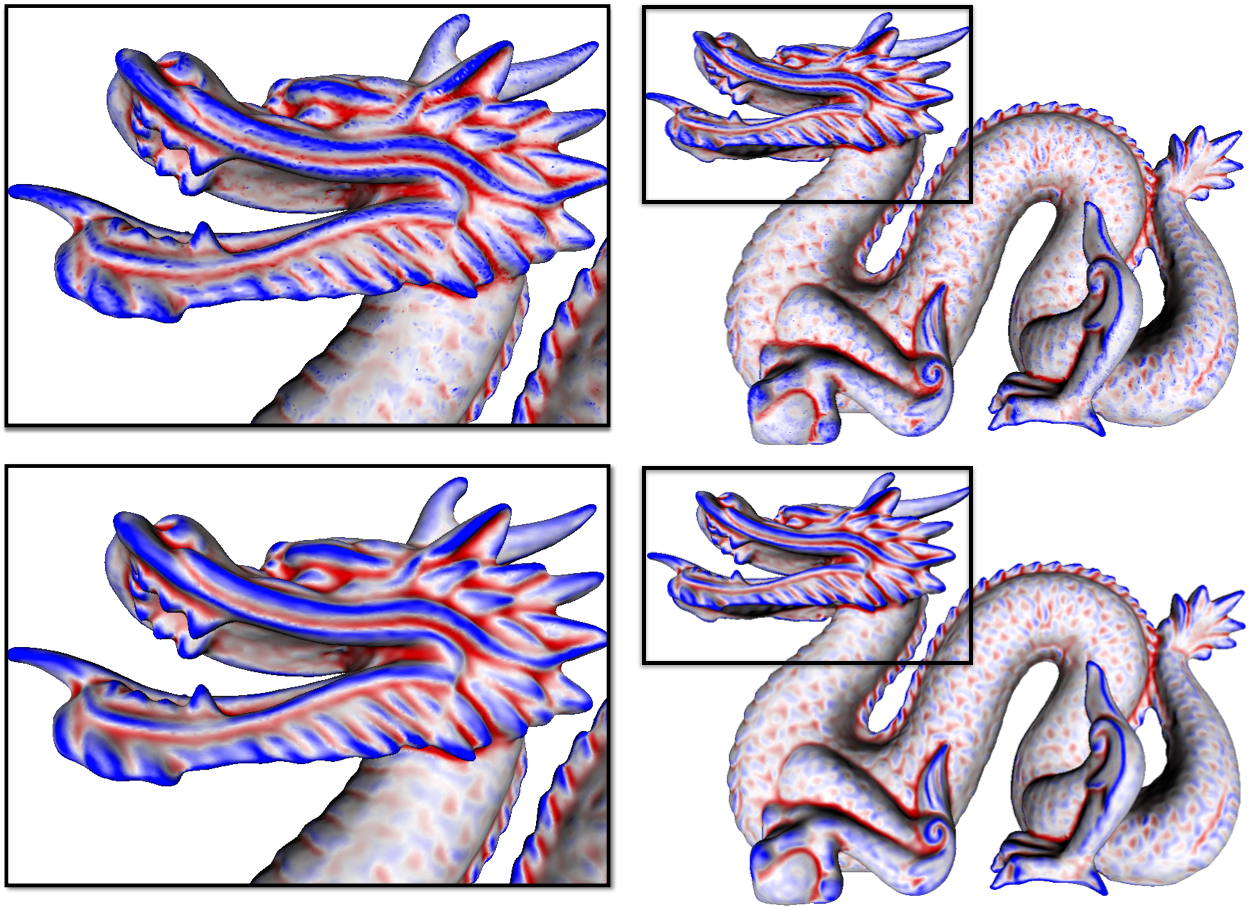}
         \caption{Dragon}
         \label{curv_dragon}
     \end{subfigure}
    \caption{Estimating mean curvature using neural implicit functions. On both images, we present the curvature at each vertex of the Bunny (\ref{curv_bunny}) and Dragon (\ref{curv_dragon}) overlaid as color on the mesh. On the top of each figure, the discrete mean curvature was calculated using the method in \citep{meyer2003discrete}, while on the bottom of each figure, we used $\text{div} \frac{\grad{f_\theta}}{\norm{\grad{f_\theta}}}$, where $f_\theta$ approximates an SDF of meshe.}
    \label{f-bunny}
\end{figure}

Another application of our work is the use of a neural network $f_\theta:\R^3\to \R$ to estimate differential properties of a triangle mesh $T$.
We train $f_\theta$ to approximate the SDF of $T$.
Since the vertices of $T$ lie in a neighborhood of the zero-level set of $f_\theta$ we use the network to \textit{map} properties of its level sets to $T$.
Afterwards, we can exploit the differentiability of $f_\theta$ to estimate curvature measures on $T$. Figure~\ref{f-bunny} shows an example of this application. We trained two neural implicit functions to approximate the SDF of the Bunny and Dragon models. We then analytically calculate the mean curvature on each vertex by evaluating $\text{div} \frac{\grad{f_\theta}}{\norm{\grad{f_\theta}}}$. Compared to classical discrete methods, the curvature calculated using $f_\theta$ is smoother and still respects the global distribution of curvature of the original mesh.
We computed the discrete mean curvatures using the method proposed by \citet{meyer2003discrete}. For our method, we used PyTorch's automatic differentiation module (\texttt{autograd})~\citep{NEURIPS2019_9015}.

\subsection{Anisotropic shading}

Another application is the use of the principal directions of curvatures in the rendering.

\begin{figure}[h!]
    \centering
        \includegraphics[width=\columnwidth]{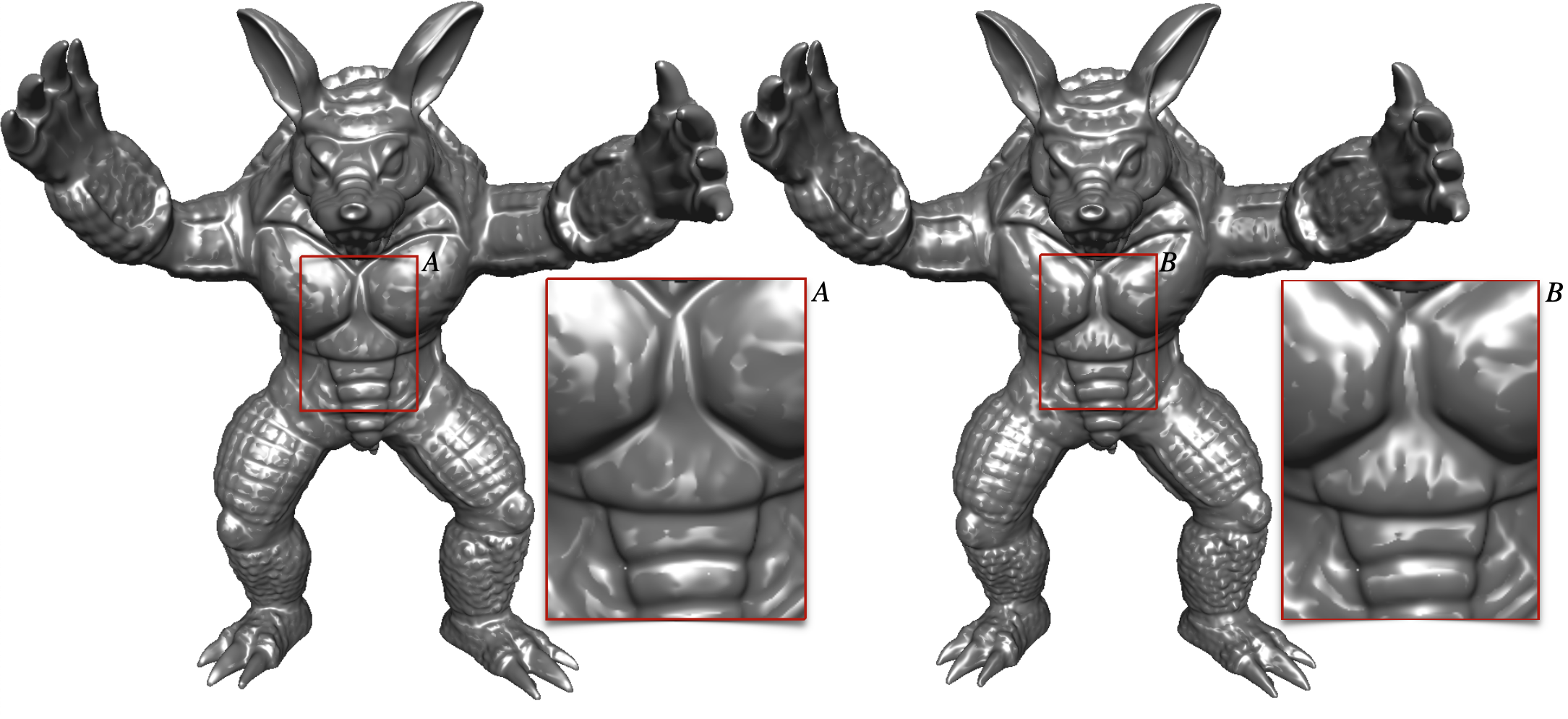}
    \caption{Ward anisotropic specular reflectance on the Armadillo neural surface. On the left, we consider a high deviation towards the maximum curvature directions. On the right, it presents the analogous for the minimum curvature directions. The principal directions were computed analytically using \texttt{PyTorch}.}
    \label{f-armadillo_ward}
\end{figure}

Let $f_\theta$ be a network such that its zero-level set $S_\theta$ approximates the Armadillo.
We present a sphere tracing visualization of $S_\theta$ using its intrinsic geometry. For this, we consider \texttt{PyTorch} to compute the shape operator of $S_\theta$.
We use its principal directions $v_1$ and $v_2$ to compute an \textit{anisotropic} shading based on the \textit{Ward reflectance model}~\citep{ward1992measuring}. It consists of using the following specular coefficient at each point $p\in S_\theta$.
$$\displaystyle k_{spec}=\frac{1}{4\pi \alpha_1 \alpha_2\sqrt{\dot{N}{l}\dot{N}{v}}}\cdot\exp\Bigg( -2\frac{\Big(\frac{\dot{H}{v_1}}{\alpha_1}\Big)^2+\Big(\frac{\dot{H}{v_2}}{\alpha_2}\Big)^2}{1+\dot{N}{H}}\Bigg).$$
Where $N$ is the normal at $p$, $v$ is the unit direction from $p$ to the observer, $l$ is the unit direction from $p$ to the light source, $H=\frac{v+l}{\norm{v+l}}$, and $\alpha_i$ are two parameters to control the anisotropy along the principal directions $v_i$.
Figure~\ref{f-armadillo_ward} presents two anisotropic shadings of $S_\theta$. The first considers $\alpha_1=0.2$ and $\alpha_2=0.5$, and the second uses $\alpha_1=0.5$ and $\alpha_2=0.2$.
We used white as the specular color and gray as the diffuse color.

\subsection{Limitations}
The main bottleneck in our work is the SDF estimation for off-surface points. We use an algorithm implemented in the \texttt{Open3D} library~\citep{zhou2018}. Even with parallelization, this method still runs on the CPU, thus greatly increasing the time needed to train our neural networks.
Also our method is designed to run with implicit visualization schemes, such as sphere tracing. However the inference time still does not allow for interactive frame-rates using traditional computer graphics pipelines. Besides recent advances in real-time visualizations of neural implicits~\citep{silva21mipplicits, takikawa2021neural}, this is still a challenge for future~works.
Finally, surfaces with sharp edges can not be accurately represented using smooth networks. 
Thus, trying to approximate them using smooth functions may lead to inconsistencies.

\subsection{Hardware}
To run all of those comparisons and tests, we used a computer with an i7-9700F with 128GiB of memory and an NVIDIA RTX 3080 with 10GiB of memory. Even to run our model in modest hardware, our strategy is lightweight with 198.657K parameters and 197.632K multiply-accumulate (MAC) operations. Running on another computer with an AMD Ryzen 7 5700G processor, 16GiB of memory, and an NVIDIA GeForce RTX 2060 with 6GiB of memory, our model took 1.32 seconds to process 172974 point samples of the Armadillo mesh.

\section{Conclusions and future works}
We introduced a neural network framework that exploits the differentiable properties of neural networks and the discrete geometry of point-sampled surfaces to represent them as neural surfaces.
The proposed loss function can consider terms with high order derivatives, such as the alignment between the principal directions.
As a result, we obtained reinforcement in the training, gaining more geometric details.
This strategy can lead to modeling applications that require curvature terms in the loss function. For example, we could choose regions of a surface and ask for an enhancement of its~features.

We also present a sampling strategy based on the discrete curvatures of the data. This allowed us to access points with more geometric information during the sampling of minibatches.
As a result, this optimization trains faster and has better geometric accuracy, since we were able to reduce the number of points in each minibatch by prioritizing the important~points.

This work emphasized the sampling of on-surface points during the training. Future work includes a sampling of off-surface points. Using the \textit{tubular neighborhood} of the surface can be a direction to improve the sampling of off-surface points.

\vspace{-0.2cm}
\section*{Acknowledgments}
We are very grateful to the anonymous reviewers for their careful and detailed comments and suggestions.
We gratefully acknowledge the support from CNPq.

\vspace{-0.2cm}

\bibliographystyle{plainnat}
\bibliography{bibliography}


\end{document}